\documentclass[useAMS,usenatbib]{mn2e}
\usepackage{epsfig}
\usepackage{times}
\usepackage{journalnames}
\usepackage{rotating}

\def\MSun{\rm {M}_{\odot}}
\def\RSun{\rm {R}_{\odot}}
\def\xitau{$\xi$\,Tau}
\def\HD{HD\,97131}

\def\apgt{\ {\raise-.5ex\hbox{$\buildrel>\over\sim$}}\ }
\def\aplt{\ {\raise-.5ex\hbox{$\buildrel<\over\sim$}}\ }

\newsavebox{\savefig}

\title[Triples with a Roche-lobe filling outer star]{The evolution of triples with a Roche-lobe filling outer star}
\author[N. de Vries, S. Portegies Zwart and J. Figueira]{
 N. de Vries,
 S. Portegies Zwart and 
 J. Figueira \\
 Leiden Observatory, Leiden University, PO Box 9513,
 2300 RA, Leiden, The Netherlands 
}

\begin{document}

\date{Accepted 2013 September 4.  Received 2013 August 21; in original form 2013 May 8}

\maketitle

\begin{abstract}
The evolution of triples has not attracted much attention in the
literature, although their evolution can be dramatically different 
from binaries and single stars. Triples are quite common, and we find 
that for about 1\% of the triples in the Tokovinin catalogue of 
multiple stellar systems in the solar neighbourhood, the tertiary star 
will overflow its Roche lobe at some time in its evolution, before any 
of the inner stars leave the main sequence. For two of these systems, 
\xitau\ and \HD, we simulate in detail this phase of mass transfer, 
during which stellar evolution, gravitational dynamics and 
hydrodynamics all play an important role. We have used the 
Astrophysical Multi-purpose Software Environment (\texttt{AMUSE}) to 
solve these physical processes in a self-consistent way. The
resulting evolution, mass transfer and the effects on the inner
as well as on the outer orbit are profound, although it is not trivial
to predict the eventual consequence of the phase of mass transfer and
the appearance of the resulting system.
\end{abstract}


\section{Introduction}

Stars tend to be formed in pairs, and many of these binaries are found
to be the member of a triple or even higher order systems.  Multiple
star systems are generally hierarchical, i.e. a binary is orbited
by another star, or another binary, etc. \citep{1983ApJ...268..319H}. 
Democratic triples (or higher order
systems) tend to be short-lived, and dissolve to lower order systems
\citep{2007MNRAS.379..111V}.

The evolution of triples has not attracted much attention in the
literature, in part due to the complications which already arise in the study of
binary systems, and which only become more severe for higher order systems.
Most studies of triples discuss the complicated
dynamical evolution without paying much attention to the stellar
aspect, whereas others focus on the stellar evolution aspects
but ignore the non-regular dynamics
\citep{1996epbs.conf..345E,1999ApJ...511..324I,2001ARep...45..620K}.
The majority of these studies focus on the formation of X-ray binaries
\citep{1986MNRAS.220P..13E}, millisecond pulsars
\citep{2011ApJ...734...55P}, and/or type Ia supernovae
\citep{2013MNRAS.tmp..704H}. 

Some studies focus on how the evolution of the triple is affected by
the non-linear gravitational dynamics of the three stars, for example
by the effect of Kozai resonances \citep{1962AJ.....67..591K}, but the
hydrodynamical effects are often reduced to semi-analytic
approximations \citep[see][]{1986MNRAS.220P..13E}.  In all cases the
presence of a third star has a major effect on the evolution of the
entire system, and the resulting product is considered rather exotic.
Triple evolution, for that reason, is often adopted to explain exotic
systems.

Triples however, are quite common \citep{2010yCat..73890925T} and 
a large fraction of stars, and therefore of binaries, are affected by
the evolution of the third companion.  In most cases these mutual
influences are minor in the sense that the third, outer star affects
the evolution on the inner binary at a level that is not reaching a
part of parameter space inaccessible by normal binary evolution. In
those cases the presence of a third star has no extraordinary effect
on the evolution of an inner binary, and regular binary evolution plus
a single star that happens to be relatively nearby can be adopted to
describe the system. In some, relatively rare, cases there is 
strong mutual interaction, and topology and orbital parameters of the system
requires a special treatment to understand its future evolution. These
cases are possibly less rare than we might think naively, and in this
paper we will discuss a seemingly rare case of triple evolution in
which gravity, stellar evolution and hydrodynamics play an important
role.

The Tokovinin catalogue of multiple stellar systems in the solar neighbourhood contains 725 triples \citep{2010yCat..73890925T}. In 130 ($\sim
18$\%) of these, the outermost star happens to be the most massive star in the system, and is therefore expected to
leave the main sequence and evolve along the giant branch before any
other star in the system. If the outer orbit is sufficiently small,
the tertiary star may overflow its Roche-lobe at some time in its
evolution. The
consequences of this rather exotic scenario are profound, and in this
paper we study the effects of Roche-lobe overflow (RLOF) of a tertiary star to an inner binary.

Upon a search for how often this scenario could possibly occur, we
found a total of six systems in the \cite{2010yCat..73890925T} catalogue that comply
to our criteria.  In these six systems we found that the semi-major axis of the
outer orbit is sufficiently small for the tertiary star to overflow
its Roche lobe at some time in its evolution, before any of the inner
stars leave the main sequence. In those cases mass will be
transported from the outer star onto the inner binary.

We simulate in detail the consequence of the onset of mass transfer for two of
these systems, \xitau\ and \HD, using a combination
of stellar evolution, gravitational dynamics and hydrodynamics.  The
resulting evolution, mass transfer and the effects on the inner
as well as on the outer orbit are profound, although it is not trivial
to predict the eventual consequence of the phase of mass transfer and
the appearance of the resulting system.

\section{Triples for which the tertiary star initiates Roche-lobe overflow first}

\begin{figure}
 \centering
 \psfig{file=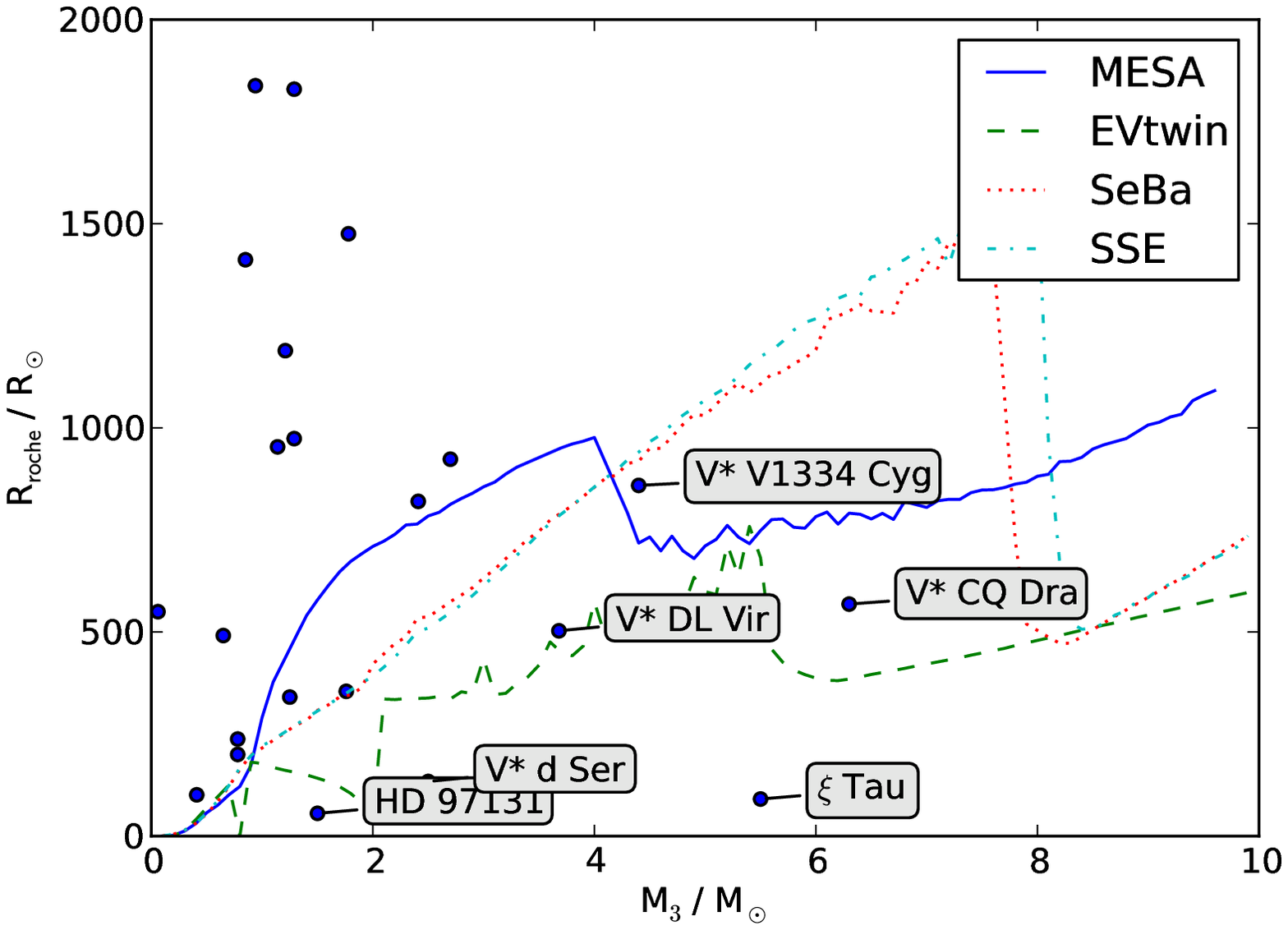,width=0.47\textwidth}
 \caption{Radius of the Roche lobe of the third component of known
   triple systems vs. the mass of the third component (bullets).  The curves
   give the maximum radius reached for a star of a certain mass according to the
   stellar evolution models \texttt{MESA} \protect\citep{2011ApJS..192....3P}, \texttt{EVtwin}
   \protect\citep{2008A&A...488.1007G, 2006epbm.book.....E, 1971MNRAS.151..351E}, \texttt{SeBa}
   \protect\citep{1996A&A...309..179P} and \texttt{SSE} \protect\citep{2000MNRAS.315..543H} for
   solar metallicity.  The triples below the curve are identified and
   their parameters are listed in Tab.\,\ref{Tab:RLOF3}.  }
 \label{fig:RLSize}
\end{figure}

Adopting the \cite{2010yCat..73890925T} catalogue of 725 multiple stellar systems in the
vicinity of the Sun, we selected those
triples for which the outer star exceeds the mass of any of the inner
stars, leaving a total of 130 triples.
In Fig.\,\ref{fig:RLSize} we show, for the outer component of these triples, the mass and estimated Roche-lobe radius \citep{1983ApJ...268..368E}, using the orbital parameters provided in the
catalogue (see also Tab.\,\ref{Tab:RLOF3}).

The curves in Fig.\,\ref{fig:RLSize} give the maximum radius as a function of initial mass, for a
series of stellar evolution calculations using a variety of codes.
The two parameterized stellar evolution codes, \texttt{SeBa} and \texttt{SSE}, give
comparable radii, but because they are mainly based on the same
evolutionary track this is not a surprise.  The smallest radii come
from \texttt{EVtwin}, which is a 1-dimensional Henyey stellar evolution code with an
extensive nuclear network. The code was run until it failed to converge, which
generally occurred at the Helium flash ($1 M_\odot \aplt M_{initial} \aplt 2 M_\odot$), thermal pulses on the asymptotic giant branch ($2 M_\odot \aplt M_{initial} \aplt 5 M_\odot$), or the Carbon flash ($M_{initial} \apgt 5 M_\odot$). \texttt{MESA} turns out to be somewhat more stable, in
particular for the relatively low mass $\aplt 4\,\MSun$ stars.  The
differences between \texttt{MESA} and \texttt{EVtwin} are quite large, which we mainly
attribute to the convergence problems of the latter.

The triples above the curves in Fig.\,\ref{fig:RLSize}
are not expected to engage in a phase of mass transfer initiated by
the outer star. For these the tertiary is expected to evolve into a
white dwarf. The final phase of these triples, however, can be quite
interesting as the extended envelope will pass the inner binary
system, much in the same way as was recently studied for planets in
orbit around post common-envelope binary systems
\citep{2013MNRAS.429L..45P}.  These systems regretfully, are beyond the
scope of our current study.

Six triples seem to
have an orbital separation sufficiently small that the outer star, at
some time after leaving the main-sequence, will overfill its
Roche lobe and transfer mass to the inner binary system.
In Fig.\,\ref{fig:RLSize} we identify those systems by their common
name. 

For further study we select two of these six systems, with relatively well-constrained orbital parameters.
These are \xitau\ and \HD, which are likely to develop Roche-lobe overflow near the end, and roughly half-way of the first ascend, respectively, both leading to type B mass
transfer \citep{1967ZA.....65..251K}.

Interestingly, the third component in these systems tends to be less 
massive than the inner binary. In this respect, Roche-lobe overflowing 
triples differ fundamentally from binaries, in which the star that 
first fills its Roche lobe is also the most massive dynamical 
component. Roche-lobe overflow in triples is therefore expected to be 
more stable as compared to binaries.

\begin{table*}
\centering
\caption{Known triples for which the outer (tertiary) star is more
  massive than any of the two inner stars and the Roche-radius is
  smaller than the maximum size of the star at the tip of the AGB.
  Columns 1 and 2 give the name and common name of the triple system; column 3 gives the record number within the \protect\cite{2010yCat..73890925T} catalogue; the numbered notes in column 4 are explained below; columns 5 to 7 give the masses of the triple components; columns 8 to 11 give the semi-major axis and eccentricity of the inner and outer orbits; column 12 gives the estimated size of the Roche lobe; columns 13 and 14 give the derived age and mass of the outer component at the onset of Roche-lobe overflow.
Remarks: (1) Masses from spectral type and therefore very uncertain; 
(2) orbits may be coplanar \protect\citep[$i_1=26^\circ$, $i_2=26^\circ$;][]{2003AJ....125..825T};
(3) possibly not even a triple \protect\citep{2003MNRAS.346..855W};
(4) Tertiary mass uncertain; 
(5) \protect\cite{2013ApJ...763...74L}.
}
  \label{Tab:RLOF3}
\begin{tabular}{lll c ccc ll ll crr}
Name & Other name & cat. &note& $M_1$ & $M_2$ & $M_3$ & $a_{\rm in}$ & $a_{\rm out}$ & $e_{\rm in}$ & $e_{\rm out}$ & $R_{\rm Roche}$ & $t_{\rm RLOF}$ & $M_{\rm RLOF}$ \\
& & no. && \multicolumn{3}{c}{[$\MSun$]} & \multicolumn{2}{c}{[AU]} & &  & [AU] & [Myr] & [$\MSun$]\\ 
\hline
HD 57061  & $\tau$ CMa   & 236 & 1 & 17.8 & 17.8 & 50.0 & 0.076 &  2.49 & 0.00    & 0.285   & 0.95 &    4.3 & 41.49 \\
HD 108907 & V* CQ Dra    & 366 & 3 &  0.8 & 0.26 &  6.3 & 0.006 &  5.42 & {\bf 0} & {\bf 0} & 2.64 &   70.1 &  6.02 \\
HD 21364  & $\xi$ Tau    &  94 & 1 &  3.2 &  3.1 &  5.5 & 0.133 &  1.23 & 0.00    & 0.15    & 0.42 &   83.7 &  5.50 \\
HD 203156 & V* V1334 Cyg & 639 &   & 1.62 & 1.83 &  4.4 & 4.616 & 10.71 & {\bf 0} & {\bf 0} & 3.99 &  167.6 &  1.67 \\
HD 120901 & V* DL Vir    & 400 & 4 &  2.5 &  1.4 & 3.68 & 0.037 &  6.68 & 0.00    & 0.46    & 2.34 &  269.4 &  3.44 \\
HD 169985 & V* d Ser     & 547 & 1 & 2.02 & 1.94 &  2.5 & 0.047 &  1.94 & 0.00    & 0.47    & 0.62 &  799.3 &  2.46 \\
HD 97131  & HD 97131     & 328 & 2 & 1.29 &  0.9 &  1.5 & 0.037 &  0.80 & 0.00    & 0.191   & 0.26 & 2959.9 &  1.48 \\

KIC002856960&  - & - &5&  0.24 & 0.22 & 0.76 &  0.008 & 0.73 &  0.0064 & 0.612 & 0.12 & $>H_0$ & 0.72 \\
\hline
\end{tabular}
\end{table*}

\section{Simulations}

\begin{figure}
    \centering
    \savebox{\savefig}{
        \psfig{file=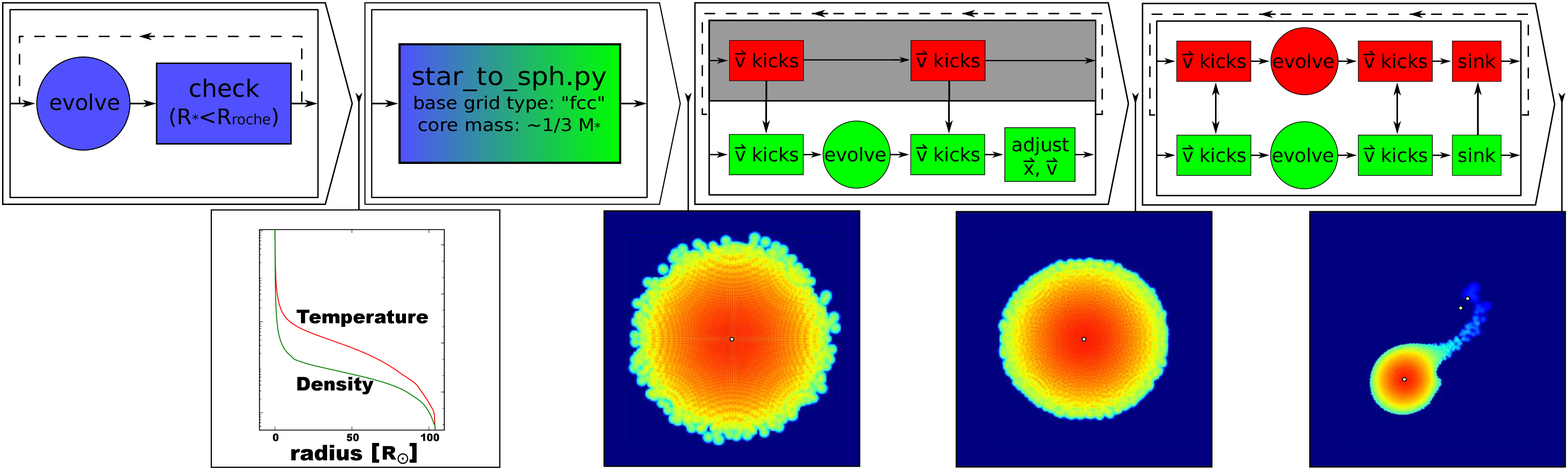, width=0.91\textheight}
    }
    \rotatebox{90}{
        \begin{minipage}{\wd\savefig}
            \usebox{\savefig}
            \caption{Schematic overview of the steps taken to simulate triples with a Roche-lobe filling outer star. Each of the four arrows describe the AMUSE solver designed for that step. Inside each combined solver, circles denote the parts performed in optimized component solvers, whereas rectangles denote operations implemented in \texttt{Python}/\texttt{AMUSE}. Colors represent the various domains of physics; red, green, and blue correspond to gravitational dynamics, hydrodynamics, and stellar evolution, respectively. After each arrow a simple plot is included to illustrate the result of the last step. Note that for the hydrodynamics plots we reduced the SPH smoothing lengths with a factor of four for enhanced contrast.}\label{fig:Schematic}
        \end{minipage}}
\end{figure}

In the evolution of outer RLOF triples, physical processes that play
an important role are stellar evolution, gravitational dynamics, and
hydrodynamics.  We have used the Astrophysical Multi-purpose Software
Environment
\cite[\texttt{AMUSE}\footnote{\texttt{http://www.amusecode.org}};][]{2009NewA...14..369P,
  2013CoPhC.183..456P} to solve all these physical processes in a
self-consistent way.  A stellar evolution code is used to model the
evolution of the outer star prior to RLOF (\S\,\ref{Sect:Sim-StellarEvolution}). At the
moment the outer star roughly fills its Roche lobe we stop the stellar
evolution simulation and convert the 1-dimensional stellar structure
from the Henyey code to a 3-dimensional hydrodynamical model (\S\,\ref{Sect:Sim-StarToSPH}). 
This hydrodynamical model of the outer star is relaxed (\S\,\ref{Sect:Sim-SPHinEquilibrium}) and put in orbit around the
binary star (\S\,\ref{Sect:Sim-Triple}).
The complex hydrodynamics of the mass
transfer from the Roche-lobe filling outer star to the inner binary is
followed for several orbits of the outer star, while keeping track of
the gravitational dynamics of the three stars and the hydrodynamics of
the gas from the outer star (\S\,\ref{Sect:Sim-CoupledEvolution}). A schematic overview of all these steps is presented in Fig.\,\ref{fig:Schematic}.
We advocate the initiative of \cite{2013arXiv1304.6780S} to require source code sharing for publication, and make all AMUSE scripts that were used for these simulations
available via the AMUSE website.

\subsection{Stellar evolution}\label{Sect:Sim-StellarEvolution}

Within the AMUSE framework we currently have the choice of four stellar
evolution codes. Two of those utilize parameterized fitting formulae to
pre-calculated stellar evolution tracks of full Henyey stellar
evolution codes. These codes can be used to acquire a quick assessment
of the estimated moment of Roche-lobe overflow of the outer star (like
we did in Fig.\,\ref{fig:RLSize}), but are not suitable for the
conversion to the hydrodynamical model of the star at the moment of
RLOF because information of the internal stellar structure is
essential for converting the stellar model to a hydrodynamical
realization.

The two Henyey stellar
evolution codes in \texttt{AMUSE} are \texttt{MESA} \citep{2011ApJS..192....3P} and \texttt{EVtwin}
\citep{2008A&A...488.1007G, 2006epbm.book.....E, 1971MNRAS.151..351E}. The stellar
structure models used in our study are generated using \texttt{EVtwin}, but there is
no particular reason for this choice over \texttt{MESA}.  All stellar
evolution calculations in this manuscript are run with the standard
AMUSE settings for these codes, adopting solar metallicity.  By
the time the outer star exceeds the radius of its Roche lobe, it already lost some mass, and its radius is considerably
larger than at birth. The calculated Roche radius $R_{\rm Roche}$, along with the age and mass at Roche-lobe overflow, $t_{\rm RLOF}$ and $M_{\rm RLOF}$, are also presented in Tab.\,\ref{Tab:RLOF3}.
In Fig.\,\ref{fig:DensityProfileVSCoreMass} the radial density profile of the outer component of \xitau\ is shown at the moment of Roche-lobe overflow (green drawn line)

\subsection{Converting the Henyey 1D stellar evolution model to a gas particles distribution}
\label{Sect:Sim-StarToSPH}

Once the radius of the outer star exceeds its Roche limit, the
1-dimensional Henyey stellar evolution model is converted to a set of
smoothed-particles hydrodynamics (SPH) particles. This is realized by 
inquiring the radial stellar structure profiles for density, 
temperature, mean molecular weight, and radius from the stellar 
evolution code. Henyey codes divide a star into a set of spherical 
shells, represented by arrays in which these parameters are stored.  
Subsequently we generate a kinematically cold set of $N$ particles 
with mass $M_{\rm RLOF}/N$ in a uniform spherical distribution. We now 
scale the particle positions radially to match the density profile of 
the star, up to its outer radius $R_{\rm RLOF}$.  Each particle is 
subsequently assigned a specific internal energy derived from the 
temperature and mean molecular mass profiles, which came from the 
stellar evolution model. This procedure is coded in the standard AMUSE 
routine called \texttt{star\_to\_sph.py}.

\begin{figure}
 \centering
 \psfig{file=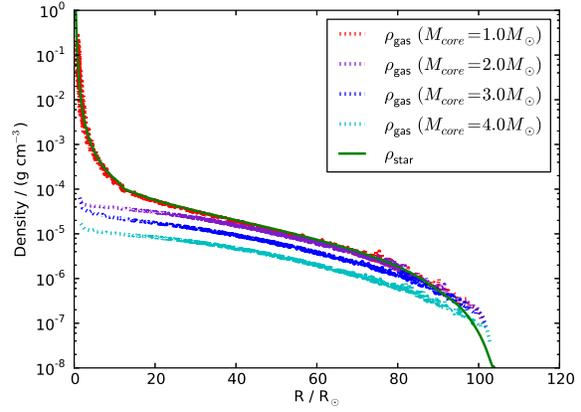, width=0.47\textwidth}
 \caption{Radial density profile of the outer component of \xitau\ at the onset of Roche-lobe overflow (green drawn line). The dotted lines represent SPH models with varying core mass $M_{\rm core}$ (red, purple, blue, and cyan dotted lines for $M_{\rm core}$\,=\,1,\,2,\,3, and 4\,$M_\odot$, respectively). {These core masses correspond to a fraction of 0.18,\,0.36,\,0.55, and 0.73 of the total mass of 5.5\,$M_\odot$.}}
 \label{fig:DensityProfileVSCoreMass}
\end{figure}

This method works very well for stellar evolution models in all phases
of the evolution. Because of the use of equal-mass particles and the
high concentration of (sub)giant stars most of the particles, and
therefore the highest resolution, will be in the stellar core, whereas
the outer edge of the star will remain barely resolved.  Most computer 
time will be spend in the stellar core due to the high density and
temperature, but in our triple evolution code, we desire to study the 
hydrodynamical effects that dominate the outer layers of the star.
These outer layers are now ill resolved, and most computer time is
spend on parts of the star that are unlikely to affect the dynamics of
the outer layers on the short hydrodynamical time scales associated
with RLOF.  Relaxing the assumption of equal mass particles is tricky
because of spurious dynamical effects associated with high-low mass
particle interactions. To prevent this, and because we are not
particularly concerned about the barely affected \citep{2005ApJ...633..418D} interior of the Roche-lobe
filling star, we replace its core with a single mass point. To prevent
the star to collapse onto itself we soften the core particle using
Plummer softening ($\epsilon$) and treat it as a pure gravitational 
point mass without pressure or internal energy. The standard cubic 
spline of \cite{1985A&A...149..135M} is used, which declines smoothly 
to zero at $2.8\epsilon$. The mass of the core particle is not
related to the mass of the hydrogen-exhausted stellar core, but a
solution to the numerical difficulties of modelling giant stars
without effecting the behaviour of the stellar envelope.  It turns out
that a core mass of about $M_{\rm core} \approx M_{\rm RLOF}/3$ gives 
satisfactory results, as illustrated for \xitau\ in 
Fig.\,\ref{fig:DensityProfileVSCoreMass} (red, purple, blue, and cyan 
dotted lines for $M_{\rm core}$\,=\,1,\,2,\,3, and 4\,$M_\odot$, 
respectively). 
{The core masses used in this figure correspond to a fraction of 0.18,\,0.36,\,0.55, and 0.73 of the total mass of $M_{\rm RLOF}$\,=\,5.5\,$M_\odot$.} 
For lower core masses, $M_{\rm core} < 0.2 M_{\rm
 RLOF}$, the problematic high density at the core remains, while for 
higher core masses, $M_{\rm core} > 0.5 M_{\rm RLOF}$, the density in 
the envelope starts to deviate appreciably from the stellar structure 
model.
{We used a core mass of 2\,$M_\odot$ for \xitau\ and 0.5\,$M_\odot$ for \HD.}

While replacing the stellar core by a single particle, we correct the 
density and internal energy profile and solve for the softening length 
of the core particle. The density and internal energy within the 
softened region are adapted in such a way that pressure equilibrium is 
maintained, while conserving the original entropy profile. The 
resulting softening length  depends on the mass of the core particle 
$M_{\rm core}$, but typically we found $\epsilon \approx 0.2 R_{\rm
 RLOF}$.
 
{Since each of our hydrodynamics simulations covers more than a thousand dynamical timescales, we adopted a moderate mass resolution with $5 \times 10^{4}$ gas particles.
We have performed shorter (5 orbits) convergence tests with up to a million particles. We noticed small variations in the orbital parameters, similar in magnitude as the temporal variations in Fig.\,\ref{fig:eXiTau} and Fig.\,\ref{fig:aXiTau}, but the general behaviour was very similar.}

\subsection{Bringing the giant in dynamic equilibrium}
\label{Sect:Sim-SPHinEquilibrium}

Even though the initial particle distribution is as smooth as 
possible, the initial particle accelerations as computed by the SPH 
code are not exactly zero. This is probably due a combination of (1) 
the stellar evolution code and the hydrodynamics code adopting a 
different equation of state, specifically using, respectively, a 
variable and a fixed adiabatic index $\gamma$, (2) numerical artefacts 
of using a (scaled) regular grid in low density regions, (3) the 
details of the implementation of gravitational softening of the SPH 
code, and (4) the treatment of particles at the surface, with 
anisotropic distributions of neighbours. If these non-zero particle 
accelerations are neglected, this will result in spurious turbulent 
velocities. Subsequently, viscous dampening will increase the internal 
energy of the particles. Therefore, some relaxation is still required 
for our models to prevent artificial heating. Since even at apocenter 
the giant is distorted from spherical symmetry, we perform the 
relaxation of the SPH model of the giant star in the potential of the 
inner binary. Performing relaxation on the giant in the gravitational presence of 
the inner binary is fairly straightforward in \texttt{AMUSE}. The 
combined solver is similar to the one that will be used for resolving 
the evolution of the triple, except for two minor changes. The 
gravitational interaction (see \S\,\ref{Sect:Sim-CoupledEvolution}) 
between the giant system (the SPH code) and the inner binary system 
(the N-body code) is evaluated one-way only: the giant system evolves 
and feels the forces of the binary, but the binary system remains 
static. Furthermore, the positions $\vec{x}_i$ and velocities 
$\vec{v}_i$ of the particles in the SPH code are adjusted after each 
step, preserving the center-of-mass position $\vec{R}_\mathrm{com}$ 
and center-of-mass velocity $\vec{V}_\mathrm{com}$, and damping the 
internal velocities by multiplying with a factor $f$ that increases 
adiabatically from 0 to 1:

\begin{equation}
 \begin{array}{r c l}
  \vec{x}_{i\mathrm{,\,adjusted}} & = & \vec{x}_i - \vec{R}_\mathrm{com} + \vec{R}_\mathrm{com, initial}\\
  \vec{v}_{i\mathrm{,\,adjusted}} & = & f \times (\vec{v}_i - \vec{V}_\mathrm{com}) + \vec{V}_\mathrm{com, initial}
 \end{array}
\end{equation}

\subsection{Setting-up the triple system}
\label{Sect:Sim-Triple}

We set up the triple from the orbital parameters taken from the
literature, which are listed in Table\,\ref{Tab:RLOF3}, and convert
those to a realization in Cartesian coordinates. The parameters that 
determine the orientation of the outer orbit relative to the inner 
orbit are the orbital inclination $i$, the longitude of the ascending 
node $\Omega$, and the argument of periastron $\omega$. Although the 
inclination of the inner and outer orbits of \xitau\ and 
\HD\ (relative to the plane of the sky) have been measured, their 
relative orbital inclination is unknown, because the longitude of 
the ascending node of the outer orbit relative to the inner orbit is 
unknown for spectroscopic triples.

Of the three parameters defining the orientation, the orbital inclination is expected to have the strongest effects on the mass transfer. 
We study these effects in \S\,\ref{Sect:Results-XiTau} for a wide range in relative orbital inclination.
For the orientation of the line of nodes we adopt the line initially connecting the two inner binary components, for the practical reason that in this case the triple configuration with the giant in hydrodynamical equilibrium is independent of the inclination, allowing us to study the effects of inclination with exactly the same initial conditions. For similar reasons we choose the argument of periastron to be $\omega=90^\circ$, since for $\omega$ close to zero the effects of the inclination probably become negligible.

For both systems the similarity of the inclination of the inner to 
that of the outer orbit are suggestive of coplanarity. Therefore we 
focus mainly on low-inclination systems in the rest of this 
paper. Note that, as the inclination approaches zero, the choice of 
longitude of the ascending node and the argument of periastron becomes 
essentially irrelevant, considering that the inner binary orbit is 
circular.
  
\begin{figure}
 \centering
 \psfig{file=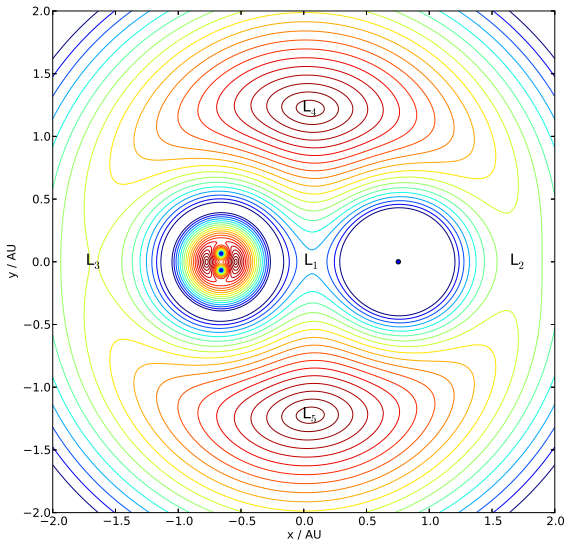, width=0.47\textwidth}
 \caption{Contour plot of the effective potential of \xitau\ in its initial configuration. The five Lagrangian points are indicated for the outer orbit. The additional set of contours, drawn inside the Roche lobe of the inner binary, also take into account the centrifugal force due to the rotation of the inner binary.}
 \label{fig:IsopotentialContourLagrangianPoints}
\end{figure}

After having constrained all initial conditions, the Cartesian
coordinates of the three stars and their masses, we can substitute the
outer star (which up to now was considered a point mass) with the 
hydrodynamical model from \S\,\ref{Sect:Sim-SPHinEquilibrium}.  The 
resulting triple is then composed of three point masses (the inner two 
stars and the central core particle of the outer giant star), and the {gas 
particles representing the stellar envelope}. A contour 
plot of the effective potential of \xitau\ in its initial 
configuration is shown in 
Fig.\,\ref{fig:IsopotentialContourLagrangianPoints}. The five 
Lagrangian points are indicated for the outer orbit. The additional 
set of contours, drawn inside the Roche lobe of the inner binary, also 
take into account the centrifugal force due to the rotation of the 
inner binary.

\subsection{Coupling hydrodynamics with gravity in the evolution model}
\label{Sect:Sim-CoupledEvolution}

We perform the simulations of RLOF in triples using a coupled 
integrator to follow the complex hydrodynamics of the mass transfer 
from the Roche-lobe filling outer star to the inner binary, while 
keeping track of the gravitational dynamics of the stars. The 
equations of motion of the inner binary is solved using the 
semi-symplectic direct N-body integrator \texttt{Huayno} 
\citep{2012NewA...17..711P}. The hydrodynamics is performed with the 
smoothed-particles hydrodynamics code \texttt{Fi} 
\citep{2005PhDT........17P}, using an adiabatic equation of state (EOS). 
{This choice of EOS was inspired by the fact that the Kelvin-Helmholtz time scale of the giant is longer than the duration of our simulations by two orders of magnitude. 
Therefore we can neglect radiative heating and cooling of the giant's gas. Cooling may become important for the gas that overflows onto the inner binary if a circumbinary disk forms. That is, however, not expected, as discussed in \S\,\ref{Sect:Theory}. 
If the Kelvin-Helmholtz time scale were very short, an isothermal EOS would be more applicable. For comparison we performed a test with an isothermal EOS. The giant model in this test was unstable, as expected. The envelope of the giant expands freely, since it is not affected by adiabatic cooling.}

{For the artificial viscosity the standard form of \cite{1992ARA&A..30..543M} is used, with $\alpha=0.5$, $\beta=1.0$, and $\eta^2=0.01$.}
Although the two inner binary stars are treated as point masses we
allow them to accrete gas from the hydrodynamics particles. This is 
realized using sink-particles which co-move with the mass points in 
the gravity code. When mass is accreted onto any of these two stars 
the masses and momenta of the particles that represent the stars in 
the N-body integrator are also adjusted properly, and the effect of 
the increased gravity of the accreting stars are properly taken into 
account while integrating their equations of motion. For the radius of 
the sink particles we adopt a value roughly twice the stellar radius, 
corresponding to $5 R_\odot$ and $2 R_\odot$ for the inner binary 
components of \xitau\ and \HD, respectively.

The N-body code as well as the hydrodynamics solver operate on their
own internal time-steps. The coupling between the two codes is
realized using the \texttt{Bridge} method in the AMUSE framework 
\citep[see Sect.\.4.3.1 in][]{2013CoPhC.183..456P}.  This coupled 
integrator is based on the splitting of the Hamiltonian much in the 
same way as is done with two different gravity solvers by 
\cite{2007PASJ...59.1095F}. The difference in our case is that one of
the two codes is an SPH code. With the adopted scheme, the
hydrodynamical solver is affected by the gravitational potential of 
its own particles, as well as the gravitational potential of the inner 
binary. The hydrodynamics, in particular the gas drag, in turn affects 
the orbits of the two inner stars. With \texttt{Bridge} we realize a 
second order coupling between the gravitational dynamics and the 
hydrodynamics. The interval at which the gravity and hydrodynamics 
interact via \texttt{Bridge} depends on the parameters of the system 
we study, but typically we achieve converged solutions when this time 
step is about a fraction of 1/64 of the inner binary orbital period.

\section{Results}
\label{Sect:CaseStudy}

Instead of simulating each of the triples in Tab.\,\ref{Tab:RLOF3} we
have selected two systems to study in more detail.  We chose
\xitau\ and \HD, without any particular reason other than
that they bracket the age range for which the outer star is likely to
fill its Roche lobe within a Hubble time and because their masses and
orbital parameters are relatively well constrained.

It is infeasible to perform simulations of the hydrodynamics of stars on a stellar evolution time scale of millions of years. We start our coupled gravity-hydrodynamics simulations when the radius of the outer star significantly exceeds its Roche radius {(with overfilling factors $R_{0}/R_{\rm Roche}$ = 1.11 and 1.07 for \xitau\ and \HD, respectively)}, to increase the mass transfer rate in order to better resolve the process of RLOF in triples. This allows us to reliably study the effects of RLOF on the orbital parameters of triples. {Mass transfer rates vary exponentially with the overfilling factor, so we will overestimate it and all consecutive rates of change of orbital parameters.} Therefore the analysis of rates of change is limited to the qualitative behaviour, but we can reliably study the relations between these rates of change.

\subsection{The case of \xitau}
\label{Sect:Results-XiTau}

The triple \xitau\ is a double-line spectroscopic binary of a
$3.2\,\MSun$ and a $3.1\,\MSun$ star.  The circular orbit has a period
of $\sim 7$\,days. The binary is orbited by a $5.5\,\MSun$ giant star in
a $\sim 144$\,day orbit with an eccentricity of 0.15
\citep{1981ApJ...246..879F}.  The outer star will start to overfill its
Roche lobe at an age of $\sim 80$\,Myr at a radius of 0.42\,AU.  The mass of
the star has by that time reduced to $5.499\,\MSun$, but we did not
correct the initial orbital separation for this small amount
($<0.1$\,\%) of mass lost in the stellar wind.  

\begin{figure}
 \centering
 \psfig{file=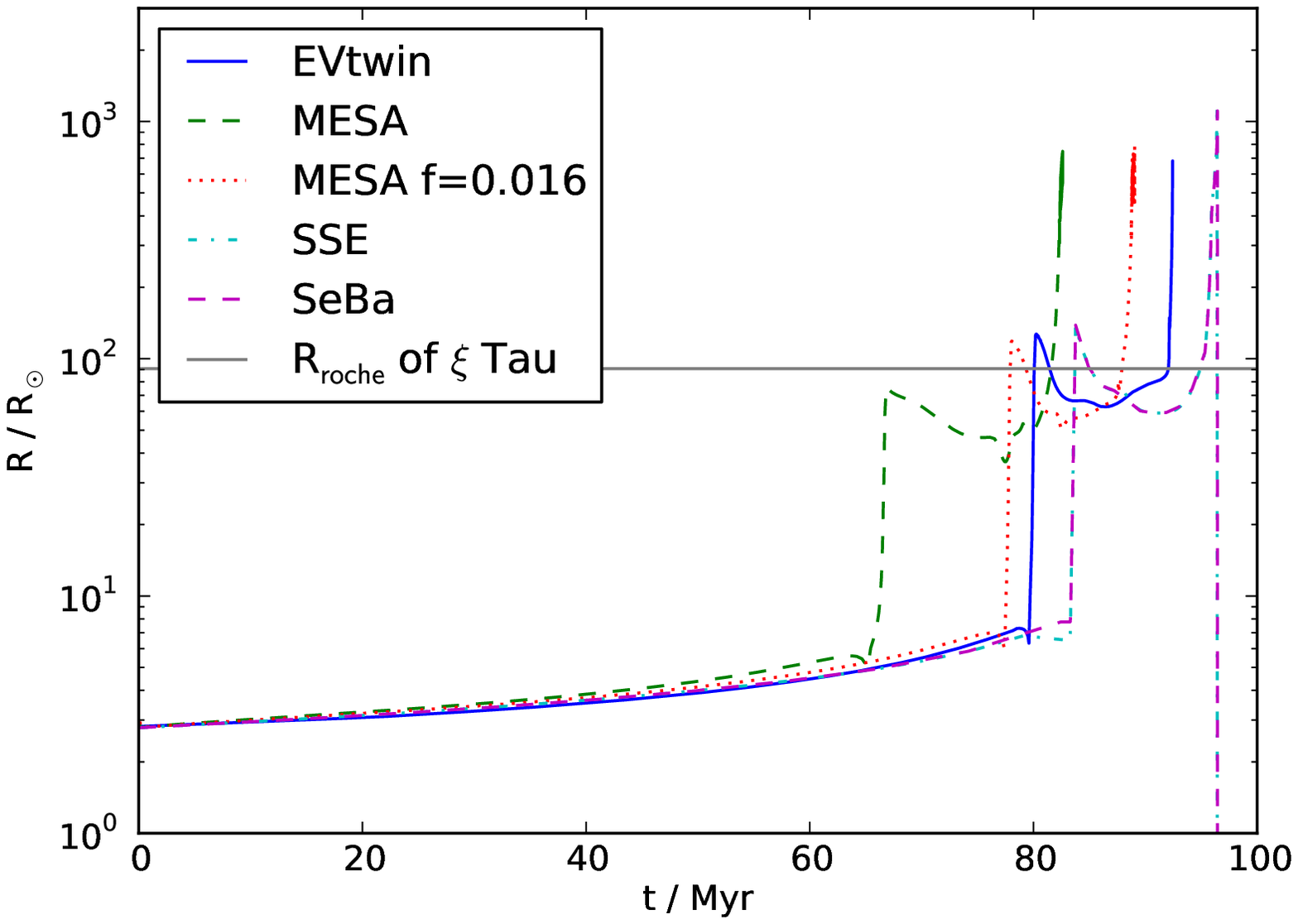,width=0.47\textwidth}
 \psfig{file=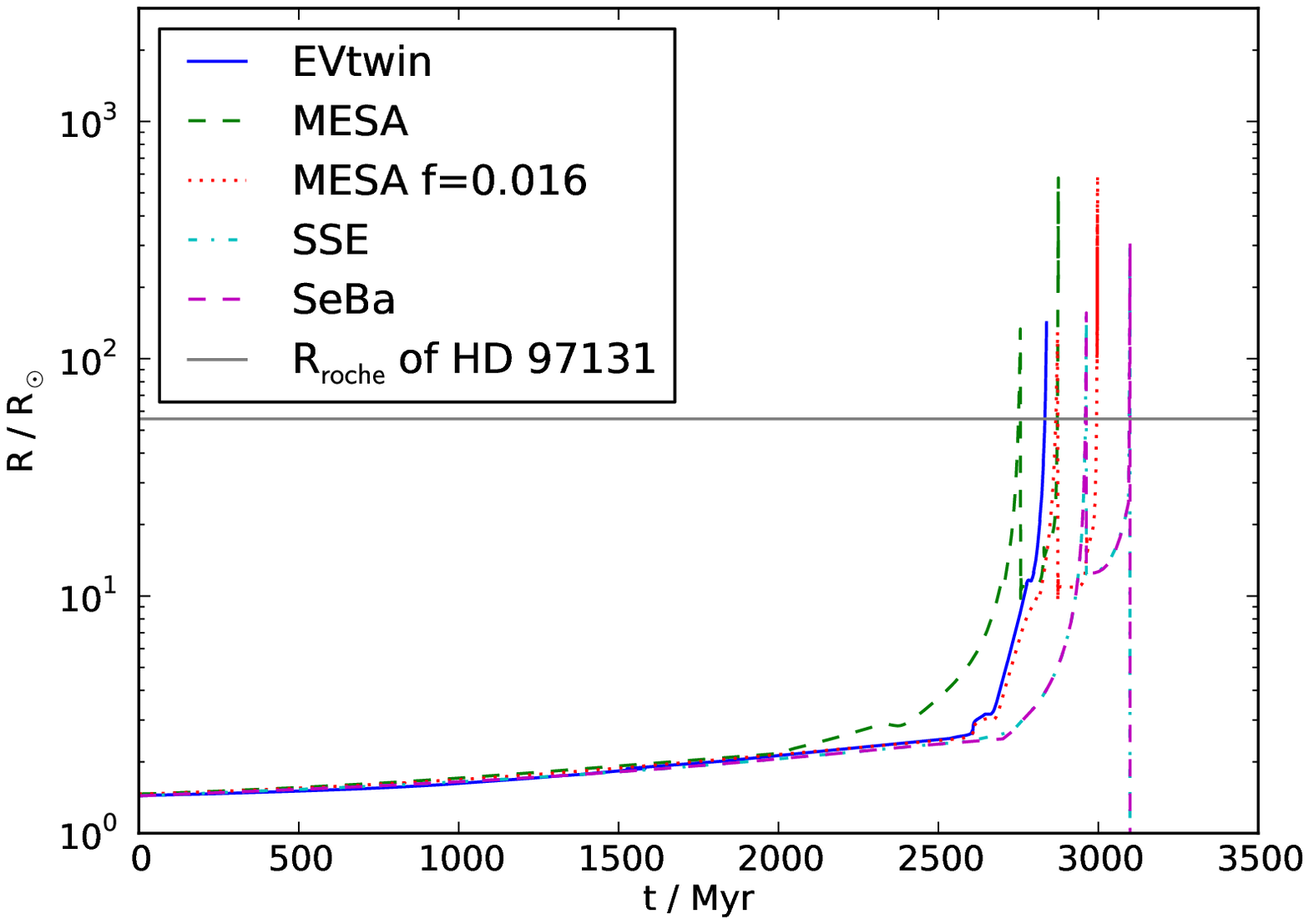,width=0.47\textwidth}
 \caption{Evolution of the tertiary star in \xitau\ (top panel) and \HD\ (bottom panel) as predicted by various stellar evolution codes: \texttt{EVtwin} (blue drawn line), \texttt{MESA} with and without convective overshoot (red dotted and green dashed lines, respectively), \texttt{SeBa} and \texttt{SSE} (magenta dashed and cyan dash-dotted lines, respectively). 
   The horizontal dotted curve gives the size of the Roche-lobe of the
   outer star in \xitau\ and \HD\ of 0.42\,AU and 0.26\,AU, respectively. }
 \label{fig:RadiusStar}
\end{figure}

In the top panel of Fig.\,\ref{fig:RadiusStar} we present the evolution of the radius of the
outer star of \xitau\ as a function of time. The various curves represent the
results calculated with different stellar evolution codes. 
The predicted evolution by \texttt{SSE} and \texttt{SeBa} virtually overlap and predict the longest lifetimes (cyan dash-dotted and magenta dashed line for \texttt{SSE} and \texttt{SeBa}, respectively).
\texttt{EVtwin} (blue drawn line) gives similar results, but \texttt{MESA} (green dashed line) predicts significantly shorter lifetimes and a smaller radius at the tip of the first giant branch.
Most codes indicate that the mode of mass transfer for \xitau\ would be consistent with case
$B$ \citep{1967ZA.....65..251K,1967AcA....17..355P}. With default parameters \texttt{MESA} predicts that  it would be case $C$, but if we include some convective overshooting (as is the default setting for \texttt{EVtwin}), \texttt{MESA} agrees on case $B$ mass transfer as well. In Fig.
\,\ref{fig:RadiusStar} we also included a simulation of the evolution using \texttt{MESA} with a typical value of $f=0.016$ \citep{2000A&A...360..952H} for the convective overshoot parameter (red dotted line).

The inclination $i$ of the orbit is constrained by the observed 
individual orbital inclinations of the inner and outer orbit, 
indicating that $i\apgt 9^\circ$ \citep{1981ApJ...246..879F}.  
However, a too high inclination would induce 
\cite{1962AJ.....67..591K} cycles. Although we do not know the current 
age of the triple, it appears unlikely to be less than one Kozai 
cycle.  If the relative inclination of the system exceeds $i \apgt 
45^{\circ}$, episodes of very high eccentricity in the inner binary 
would have had occurred, which would have initiated tidal effects,
mass transfer, or even a merger. We cannot formally rule out 
that some mass transfer may have occurred in the past of the inner binary, 
but we consider it highly unlikely that $i \aplt 45^{\circ}$. We
performed several simulations with inclination ranging from 
$i=9^\circ$ to $i=69^\circ$.

\begin{figure}
 \centering
 \psfig{file=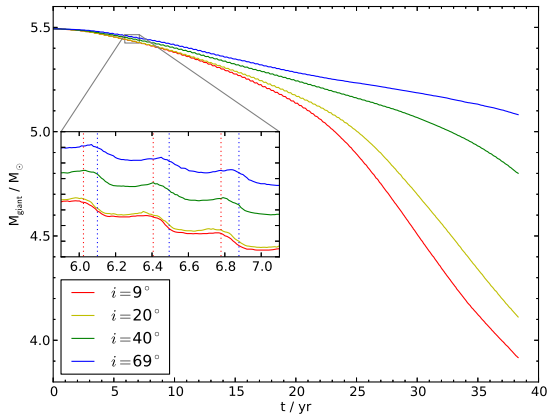,width=0.47\textwidth}
 \caption{Mass of the donor in \xitau\ as a function of time, starting
   at the onset of RLOF, for a range in initial relative inclination: $i=9^\circ$ (red), $i=20^\circ$ (yellow), $i=40^\circ$ (green), and $i=69^\circ$ (blue).}
 \label{fig:MXiTau}
\end{figure}

With the adopted parameters the outer star will fill its Roche lobe
while the other (inner binary stars) still reside on the main sequence
and are well detached from their respective Roche lobes.  In
Fig.\,\ref{fig:MXiTau} we present the evolution of the mass of the
outer star in four calculations for $i=9^\circ$, $i=20^\circ$, $i=40^\circ$, and
$i=69^\circ$. There is a clear trend, in which high inclination gives 
a considerably lower rate of mass loss, and for a lower inclination 
the rate of mass lost is relatively high. This is expected, because 
each of the inner binary stars can approach the giant more closely for 
low inclination where the orbits lie roughly in the same plane.

The mass loss from the outer giant is periodic, modulated with the 
periodicity of the outer orbit. This is a result of the slight 
eccentricity of the outer orbit, which causes the giant to overfill 
its Roche lobe at pericenter, whereas after semi-latus rectum the 
star detaches from its Roche lobe, to re-establish RLOF when it again 
approaches pericenter. This periodicity is visible in the inset of Fig.\,\ref{fig:MXiTau}, which zooms in on the time interval between 5.9 and 7.1\,years. The moments of pericenter crossing are indicated for the $i=9^\circ$ and $i=69^\circ$ simulations (dotted vertical lines with matching colors). In agreement with a study of RLOF in eccentric binaries \citep{2011ApJ...726...67L}, there is a substantial delay between the moments of pericenter crossing and the peaks in mass transfer.

\begin{figure}
 \centering
 \psfig{file=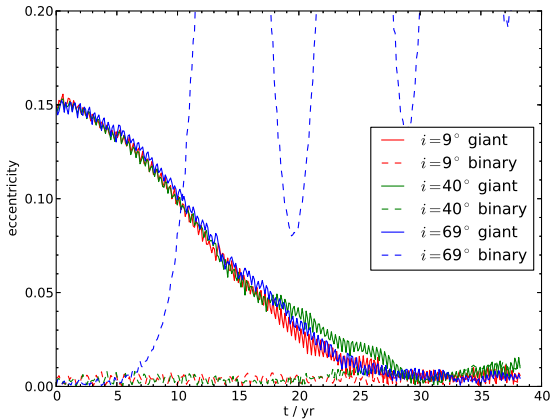,width=0.47\textwidth}
 \caption{Eccentricity of the outer (drawn lines) and inner (dashed lines) orbit in \xitau\ as a function of time, for an initial relative inclination of $i=9^\circ$ (red), $i=40^\circ$ (green), and $i=69^\circ$ (blue).}
 \label{fig:eXiTau}
\end{figure}

\begin{figure}
 \centering
 \psfig{file=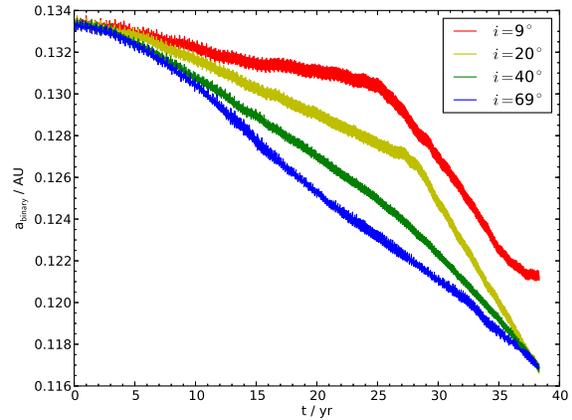,width=0.47\textwidth}
 \psfig{file=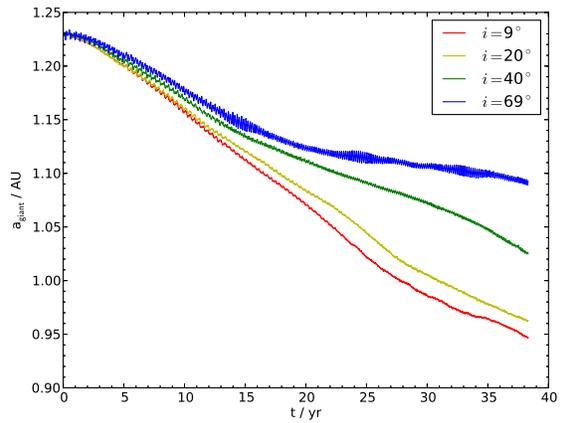,width=0.47\textwidth}
 \caption{Semi-major axis of the inner (top panel) and outer (bottom panel) orbit in \xitau\ as a function of time, for an initial relative inclination of $i=9^\circ$ (red), $i=20^\circ$ (yellow), $i=40^\circ$ (green), and $i=69^\circ$ (blue).}
 \label{fig:aXiTau}
\end{figure}

In Fig.\,\ref{fig:eXiTau} we present the evolution of the eccentricity
of the inner and outer orbits of \xitau, starting from the moment of 
RLOF. In Fig.\,\ref{fig:aXiTau} we present the evolution of the
semi-major axis. The high inclination ($i=69^\circ$) orbit evolves
towards a higher inner eccentricity, consistent with the Kozai 
effect. In itself it is interesting that our combined
gravity-hydrodynamics code in \texttt{AMUSE} reproduces the Kozai 
effect accurately. We checked this by integrating the 3-point-mass
simulations using \texttt{Huayno}, which gives a very similar 
evolution of the eccentricity of the inner orbit. The rate of mass 
loss from the giant is lower in the high inclination orbit. Therefore 
the semi-major axis of the outer orbit decreases slower in the highly 
inclined orbit than in the low-inclination orbit. As a consequence the 
rate of mass loss from the giant becomes even smaller for high 
inclinations (see Fig.\,\ref{fig:MXiTau} and Fig.\,\ref{fig:aXiTau}).  
The decay in the eccentricity of the outer orbit (see 
Fig.\,\ref{fig:eXiTau}) is independent of inclination, and is caused 
by tidal effects in the interaction between the outer giant and the 
inner binary. The effect of the hydrodynamics on the eccentricity of 
the inner orbit is negligible.  The short periodicity noticeable in 
Fig.\,\ref{fig:eXiTau} and \ref{fig:aXiTau} is a reflection of the 
motion of the inner binary stars.

\begin{figure*}
 \centering
 \psfig{file=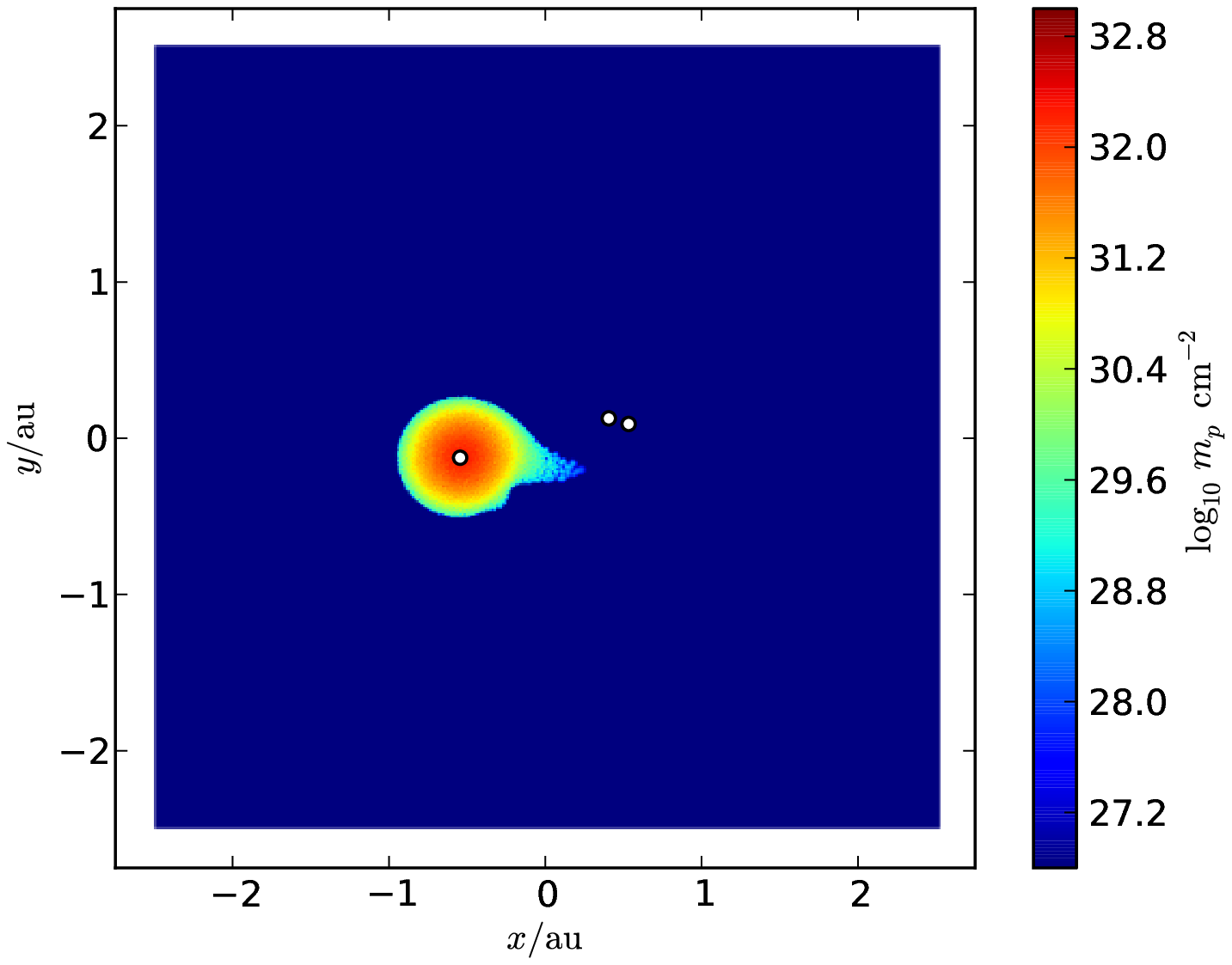,width=0.47\textwidth}
 \psfig{file=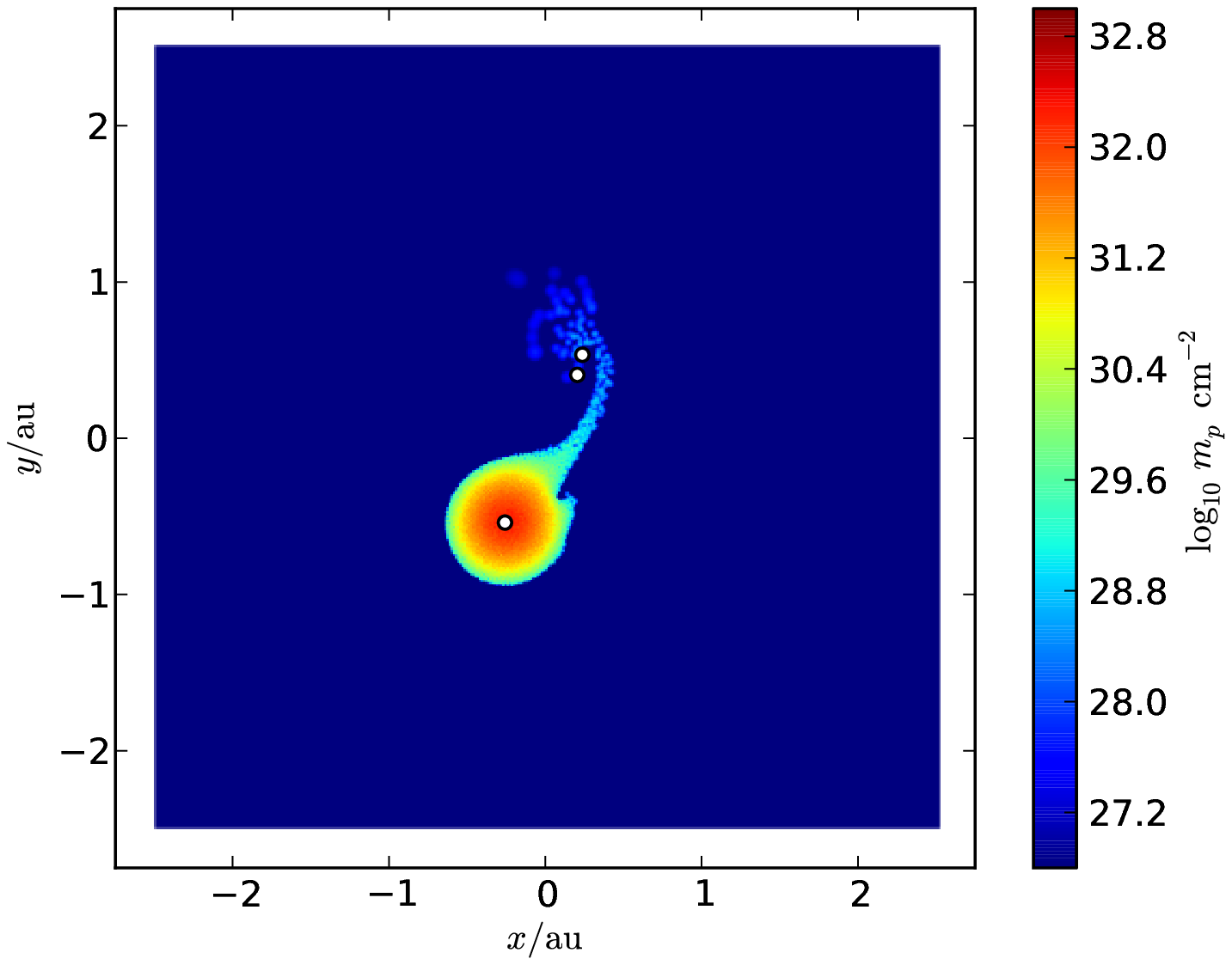,width=0.47\textwidth}
 \psfig{file=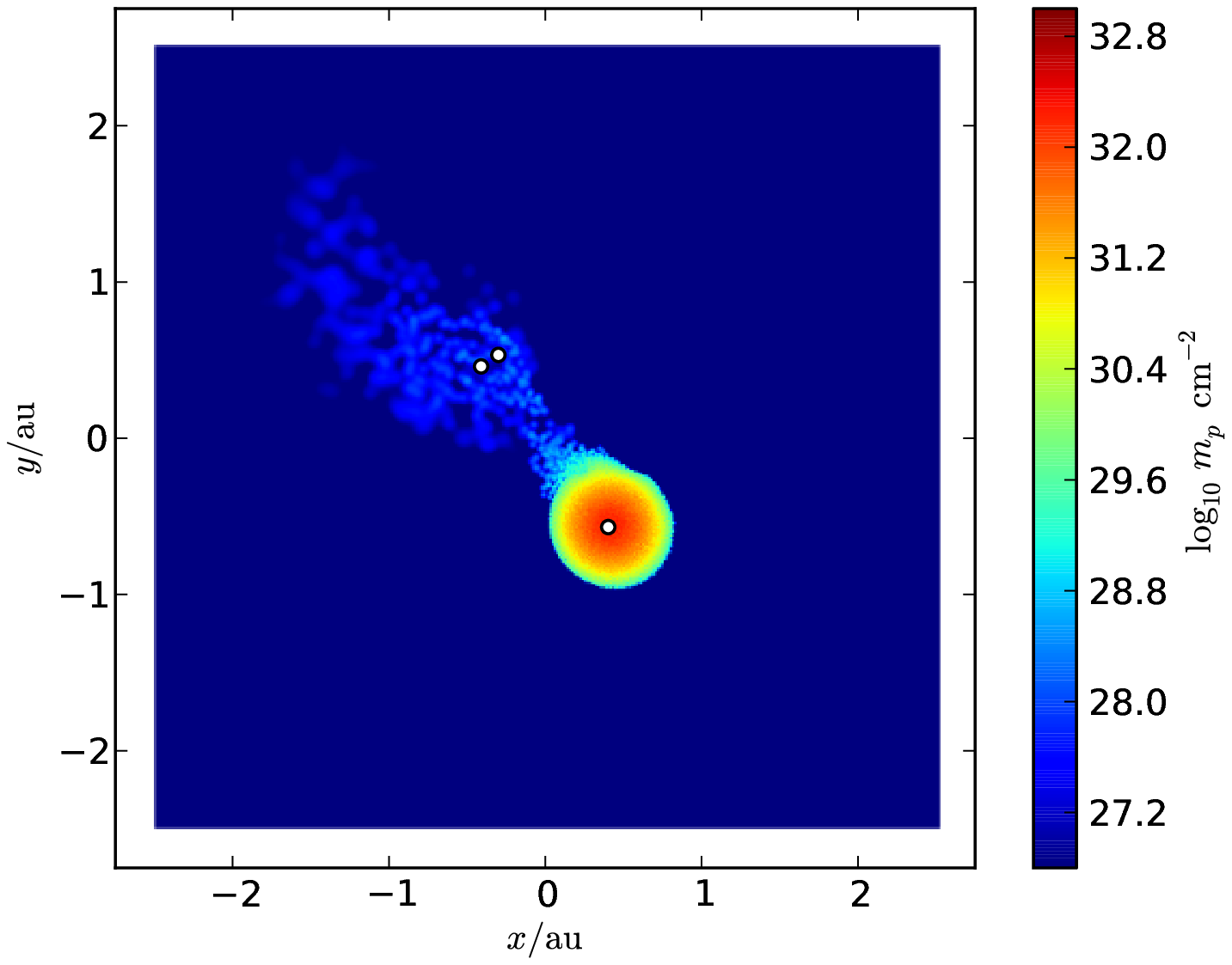,width=0.47\textwidth}
 \psfig{file=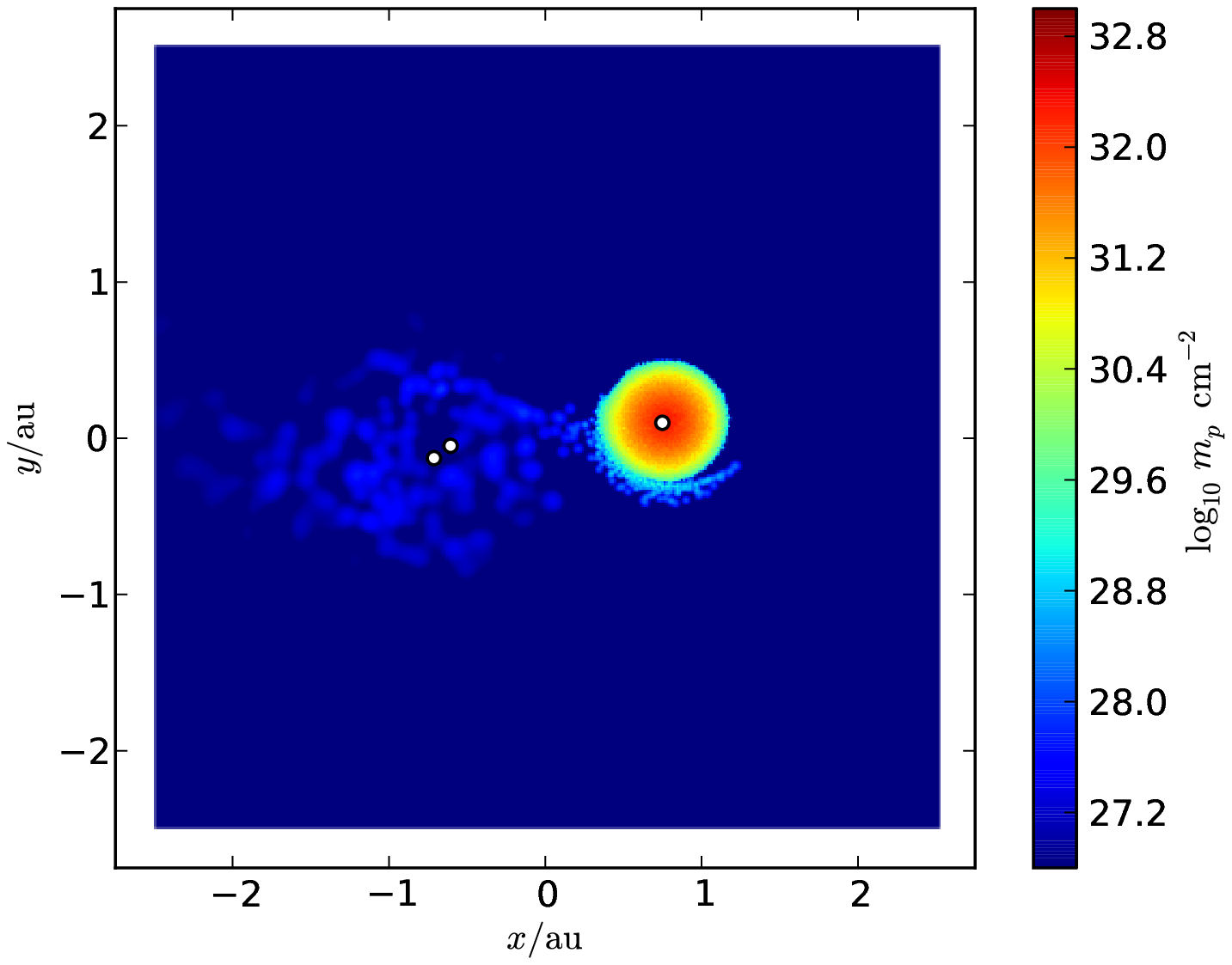,width=0.47\textwidth}
 \psfig{file=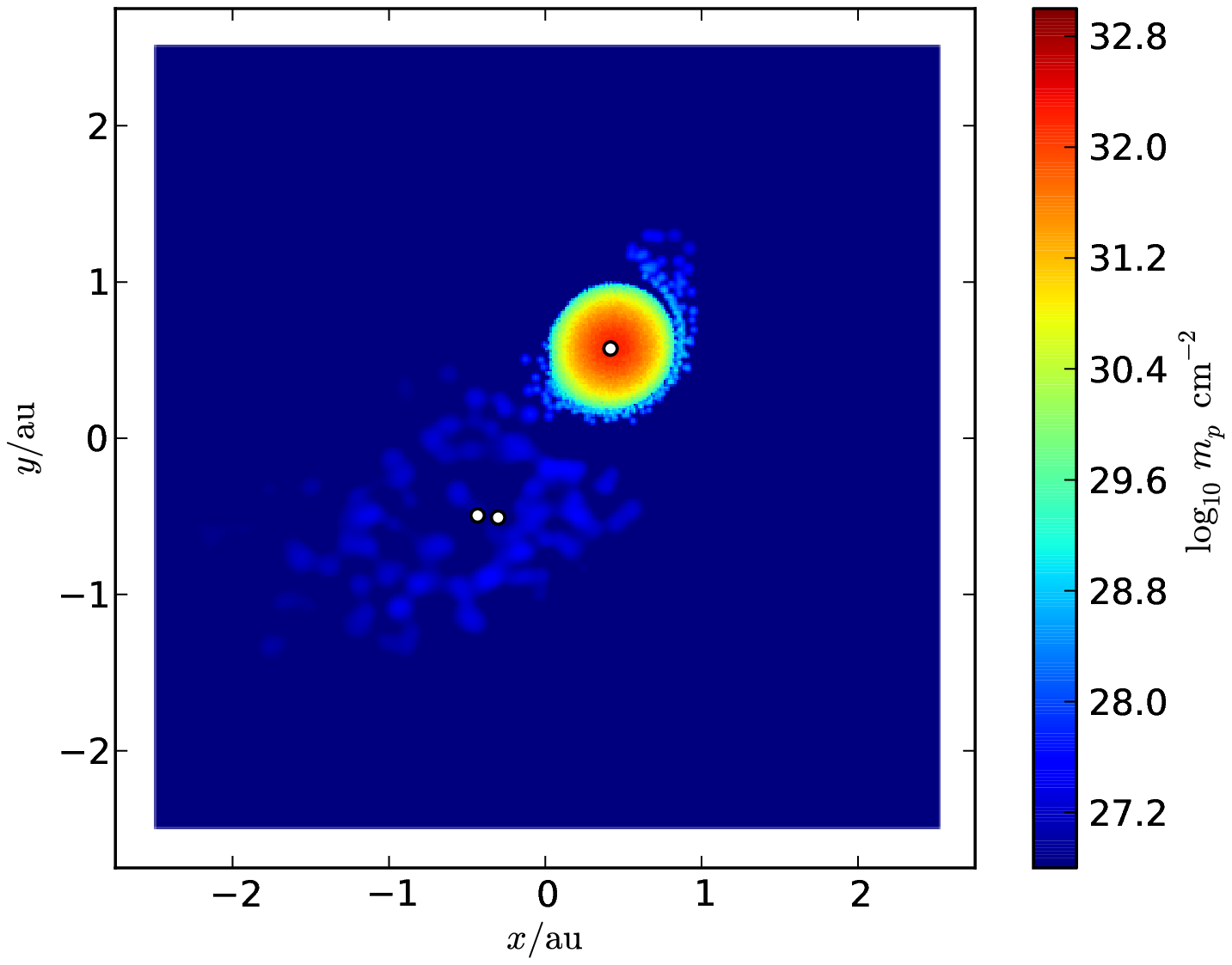,width=0.47\textwidth}
 \psfig{file=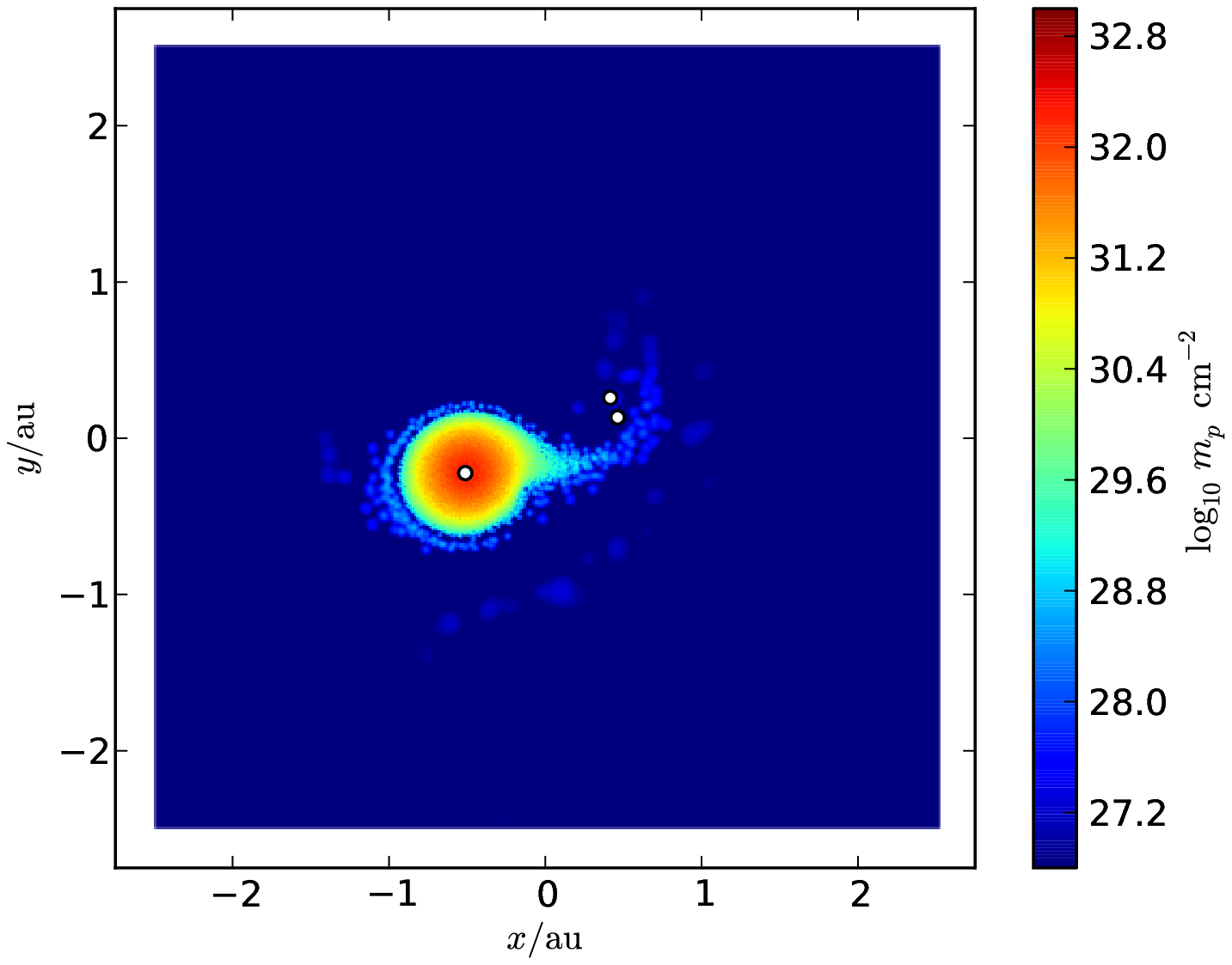,width=0.47\textwidth}
 \caption{Column density images of \xitau\ after 76, 92, 116, 148, 172, and 224 days (from left to right and from top to bottom) since the onset of RLOF.}
 \label{fig:StillsXiTau}
\end{figure*}

\subsection{The case of \HD}

The triple \HD\ is a system in which a $1.5\,\MSun$ star orbits a 
compact ($a \approx 8 \RSun$) binary of $1.29\,\MSun$ and
$0.90\,\MSun$, with a 
$\sim$134\,day period and an eccentricity of 0.191 (see Tab.\,\ref{Tab:RLOF3}).  The outer star will
start to overfill its Roche lobe at an age of $\sim$3\,Gyr at a radius of
0.26\,AU, which is consistent with mass transfer case $B$
\citep{1967ZA.....65..251K, 1967AcA....17..355P}.  The mass of the star has by that
time reduced to $1.48\,\MSun$, but like in the case of \xitau\ we did
not correct the initial orbital separation for the small amount of mass lost in the
stellar wind. 
Like is the case for \xitau, the outer star will fill its Roche-lobe
while the two inner binary stars are still on the main-sequence
and are well detached from their respective Roche lobes.  

{We agree with \cite{2003AJ....125..825T} that the inclination of the inner and outer orbit ($i_1=26^\circ$, $i_2=26^\circ$) are suggestive of coplanarity. Therefore we only performed a simulation with relative inclination $i=0^\circ$ for \HD.}

\begin{figure}
 \centering
 \psfig{file=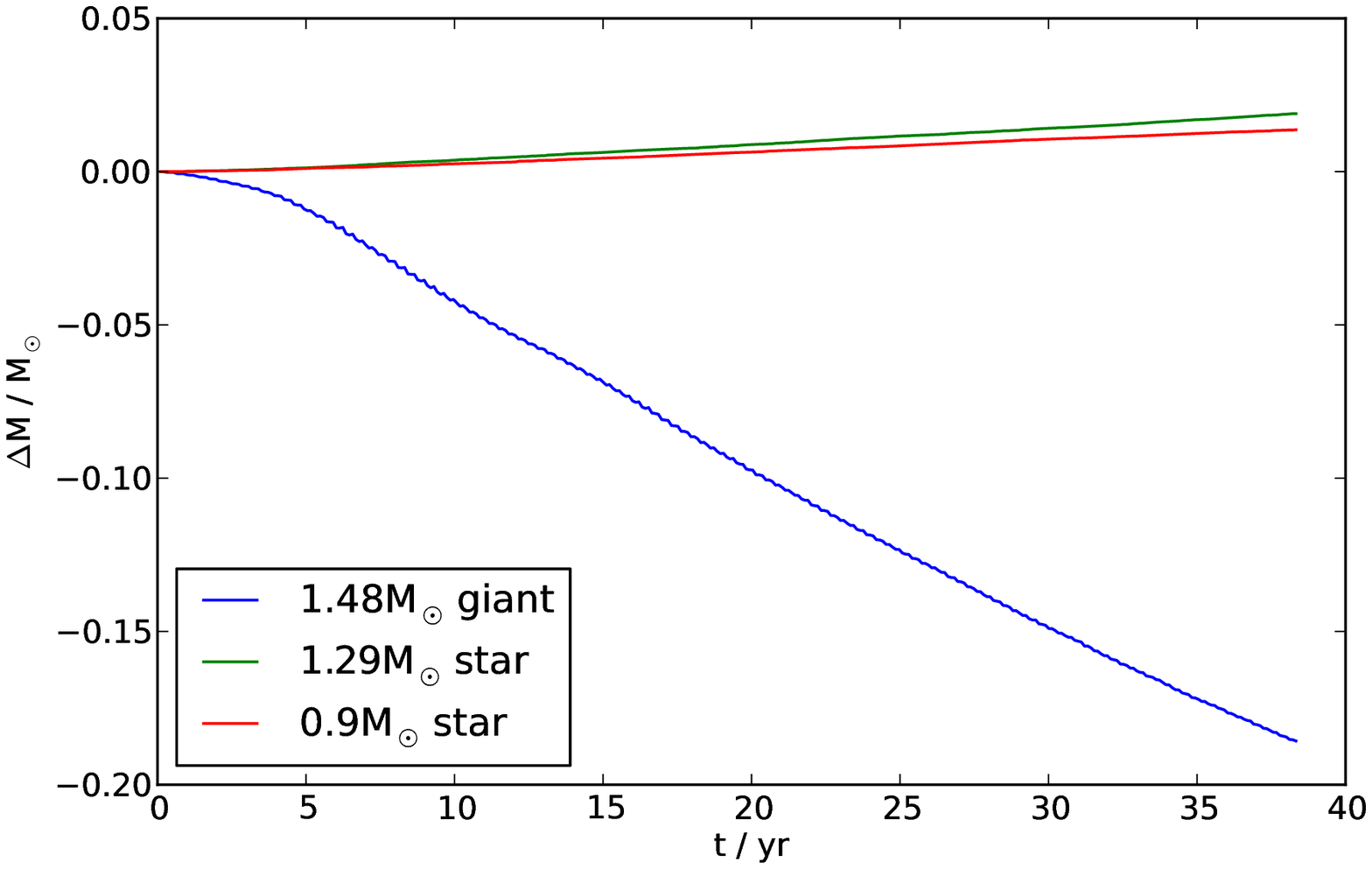,width=0.47\textwidth}
 \caption{Mass of each component of \HD, as a function of time, starting
   at the onset of RLOF.  }
 \label{fig:MHD97131}
\end{figure}

\begin{figure}
 \centering
 \psfig{file=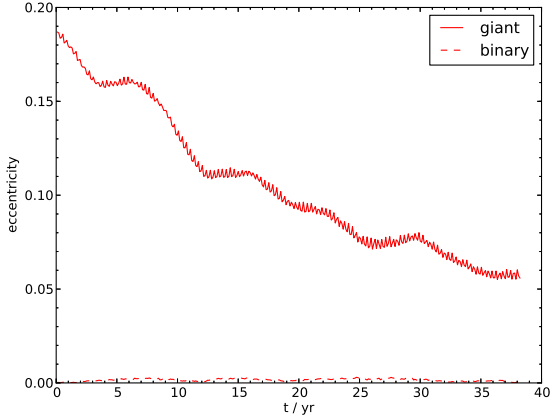,width=0.47\textwidth}
 \caption{Eccentricity evolution of the inner (dashed line) and outer (drawn line) orbit of \HD.}
 \label{fig:eHD97131}
\end{figure}

\begin{figure}
 \centering
 \psfig{file=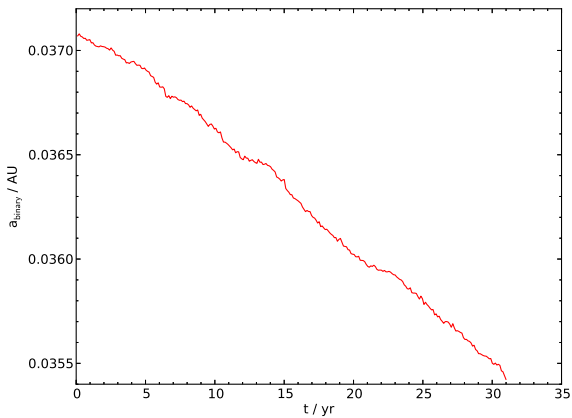,width=0.47\textwidth}
 \caption{Semi-major axis evolution of the inner orbit of \HD.}
 \label{fig:aHD97131}
\end{figure}

In Fig.\,\ref{fig:MHD97131} we present the evolution of the masses of
the three stars. 
The $1.29\,\MSun$ primary (green line) and $0.9\,\MSun$ secondary (red line) accrete only a small fraction of the mass lost by the tertiary (blue line).
After about 5\,years (14 orbits) the mass-transfer rate becomes fairly constant, while the eccentricity of the outer orbit slowly declines (Fig.\,\ref{fig:eHD97131}).
Both the companion stars accrete about 1\% of their
own mass from the giant's envelope. The less massive inner star
accretes a slightly higher fraction, but in absolute numbers the accretion rate
is smaller than that of its more massive companion. The outer star lost $0.19\,\MSun$ (13\%) of its mass, while the semi-major axes of the inner and outer orbits decreased by $0.42\,\RSun$ (5\%, see Fig.\,\ref{fig:aHD97131}) and $25\,\RSun$ (15\%, see black solid line in Fig.\,\ref{fig:OrbitalEvolution}), respectively.

\section{Theory}\label{Sect:Theory}

Based on our hydrodynamical and gravitational simulations we can
measure the evolution of the orbital parameters and the masses of all
three stars. This enables us to constrain some of the fundamental
parameters in binary and triple evolution.

To keep the model simple we assume that the mass that is lost from the
donor is either accreted onto the inner binary (either the primary
star or the secondary) or it is lost from the triple system with a
specific amount of angular momentum per unit mass.  Upon inspection of
our simulation results, no disk forms in either case, but all mass lost
by the donor star escapes the triple via the $L_2$ and $L_3$
Lagrangian points of the outer orbit. The vast majority of mass lost
by the outer star, however, is funnelled through the $L_1$ Lagrangian
point to be ejected by the inner binary system through the $L_3$
Lagrangian point of the outer orbit. We did not see a circumbinary
disk forming, which
{could be due to the limited resolution near the binary, or the fact that cooling is neglected. However, we don't even expect a circumbinary disk because of} the relatively large
orbital separation of the inner binary (See Fig.\,\ref{fig:IsopotentialContourLagrangianPoints}). 

\subsection{The evolution of the inner orbit}

The mass that flows through the first Lagrangian point of the outer
orbit onto the inner binary system can be considered as a gaseous
cloud around the binary. The accretion stream intersects with the
orbit of the inner binary and this redistributes the mass, angular
momentum and energy of the accretion stream.  We estimate the minimal
distance between the center of mass of the inner binary and the first
periastron passage of the accretion stream using the fitting formula by
\cite{1975ApJ...198..383L}. Following this prescription the accretion
stream of the following systems: HD\,57061, HD\,108907, HD\,21364 (\xitau), and HD\,203156, intersects with the orbital trajectories of both
stars of the inner binary. 
As a consequence no disk forms in these systems and the mass dumped from the outer
(giant) star onto the inner binary can be considered as some sort of a
common envelope, in which the ejection of the matter coming from the
outer star is ejected from the inner orbit by its orbital energy. This
picture is inspired by the fact that in our
calculations the majority of the mass that is provided by the giant is
ejected from the inner binary. The amount of giant's material that is
accreted by any of the inner stars is fairly small.

We can estimate the binding energy $E_b$ of the material $dM$ that enters the inner
binary of mass $m_{\rm in} = m_1 + m_2$:
\begin{equation}
  E_b = - {G (m_1 + m_2) dM \over \lambda a_{\rm in}}.
\label{Eq:bindingEnergy}\end{equation}
Here $\lambda$ is some structural parameter that describes the amount
of binding energy in the mass that is supplied by the giant. The orbital energy lost due to the ejection of the giant's
material is
\begin{equation}
  E_{{\rm orb}, i} - E_{{\rm orb}, f} = {G m_1 m_2 \over 2a_i} 
  - {G(m_1+dm_1) (m_2+dm_2) \over 2a_f} 
\label{Eq:EnergyChange}\end{equation}
Like in the classical common envelope evolution we can now compare the
binding energy of the envelope with the change in binding energy and
derive a value of the ratio in energies and the efficiency factor
$\alpha$. Because we do not really know the structure of the common
envelope in this case, we keep the two parameters together as $\alpha
\lambda$, like is proposed by \cite{1996A&A...309..179P}.

\begin{table}
\centering
\caption{Averaged values of the common-envelope efficiency parameter $\alpha \lambda$ over about 100 orbits. The error estimates are derived from the temporal variations. Column 1 gives the name of the simulated triple; column 2 the relative inclination; column 3 the measured value of $\alpha \lambda$.}
  \label{Tab:alphalambda}
\begin{tabular}{ccc}
Object name & $i$ &  $\alpha \lambda$ \\
\hline
 \xitau &  $9^\circ$ & $5.8 \pm 0.8$ \\
 \xitau & $20^\circ$ & $5.0 \pm 0.6$ \\
 \xitau & $40^\circ$ & $3.7 \pm 0.3$ \\
 \xitau & $69^\circ$ & $2.9 \pm 0.3$ \\
 \HD    &  $0^\circ$ & $6.8 \pm 0.9$ \\
\hline
\end{tabular}
\end{table}

{From the hydrodynamical calculations we now measure all parameters in
Eqs.\,\ref{Eq:bindingEnergy}\,and\,\ref{Eq:EnergyChange}. 
The results are presented in Table\,\ref{Tab:alphalambda}, with the name of the simulated triple in column 1, the relative orbital inclination in column 2, and the measured value of $\alpha \lambda$ in column 3. The presented values of the common-envelope efficiency parameter $\alpha \lambda$ are averages over about 100 orbits. The error estimates are derived from the temporal variations.
The consistency of the result from \HD\ and the low inclination simulation of \xitau\ is
remarkable, in particular because both systems have quite different
parameters.
For our simulations with high relative inclination, the measurements of $\alpha \lambda$ agree with the value derived for an observed post-common-envelope binary with a moderate mass giant \citep{2012ApJ...759L..34C}, and with the value found by high-precision measurements using planets in orbit around a post-common-envelope binary \citep{2013MNRAS.429L..45P}.
The simulations with lower relative inclination result in higher values for $\alpha \lambda$, which means that it takes less orbital energy from the inner binary to expel the gas. 
Indeed, when the angular momentum vectors of the inner binary and the inflowing gas are (almost) aligned, less angular momentum needs to be transferred to speed up the gas to the escape velocity.
}

\subsection{The evolution of the outer orbit}

For low inclination orbits, the evolution of the orbital separation
for \xitau\ and \HD\ is driven by the mass loss of the donor
star during Roche-lobe overflow.  During Roche-lobe overflow mass
leaving the outer donor star will either be accreted onto the binary
stars, form a disk around them, or is ejected from the triple system
altogether. The orbit will be affected depending on what actually
happens to the mass. We can express this in the amount of angular
momentum that is carried with the mass that leaves the donor star.

We derive the expression for the effect of mass transfer in the system
starting from a binary with primary mass $m^0_{\rm g}$ and secondary
mass $m^0_{\rm in}$ to the situation after mass transfer with masses
$m_{\rm g}$ and $m_{\rm in}$ for the primary and secondary stars,
respectively.  The degree of conservatism is then expressed in the
parameter $f = (m_{\rm g}+m_{\rm in})/(m^0_{\rm g}+m^0_{\rm in})$.
Ignoring internal angular momentum of the outer giant star and the
inner binary system, we can express the evolution of the semi-major
axis according to the redistribution of mass, which can be expressed
as \citep{1990A&A...236..107H}:
\begin{equation}
        {{\dot a} \over a} = -2{{\dot m}_{\rm g} \over m_{\rm g}}
        \times \left(1 - \beta {m_{\rm g} \over m_{\rm 2}}
        - {1 \over 2} {(1-\beta)m_{\rm g} \over m_{\rm g}+m_{\rm 2}}
        -f {(1-\beta)m_{\rm g} \over m_{\rm g}+m_{\rm 2}} \right)
\label{Eq:adotovera}\end{equation}
Here $\beta$ is the specific angular momentum of the mass that leaves
the system as a fraction of the specific angular momentum of the
accreting system (in this case the center of mass of the inner binary
system).  Integrating Eq.\,\ref{Eq:adotovera}, assuming $\beta$
and the fraction of mass lost $f$ are constant,
gives \citep{1995A&A...296..691P}:
\begin{equation}
        {a \over a_0} = \left( {m_{\rm in} m_{\rm g} \over
                                m_{\rm in}^0 m_{\rm g}^0} \right)^{-2}
        f^{2\beta + 1}
\label{Eq:NewSemi}
\end{equation}
This equation reduces to the classical conservative case
\begin{equation}
        {a \over a_0} = \left( {m_{\rm g} m_{\rm in} \over m_{\rm
        in}^0 m_{\rm g}^0} \right)^{-2}
\end{equation}
when we adopt $f=1$, in which case the evolution of the orbital
separation becomes independent of $\beta$.

The specific angular momentum of the mass that leaves the system is
uncertain. We can measure the amount of angular momentum lost from 
the system by applying conservation of angular momentum in the 
non-conservative case, like for binaries was discussed by 
\cite{1995A&A...296..691P}.

\begin{figure}
 \centering
 \psfig{file=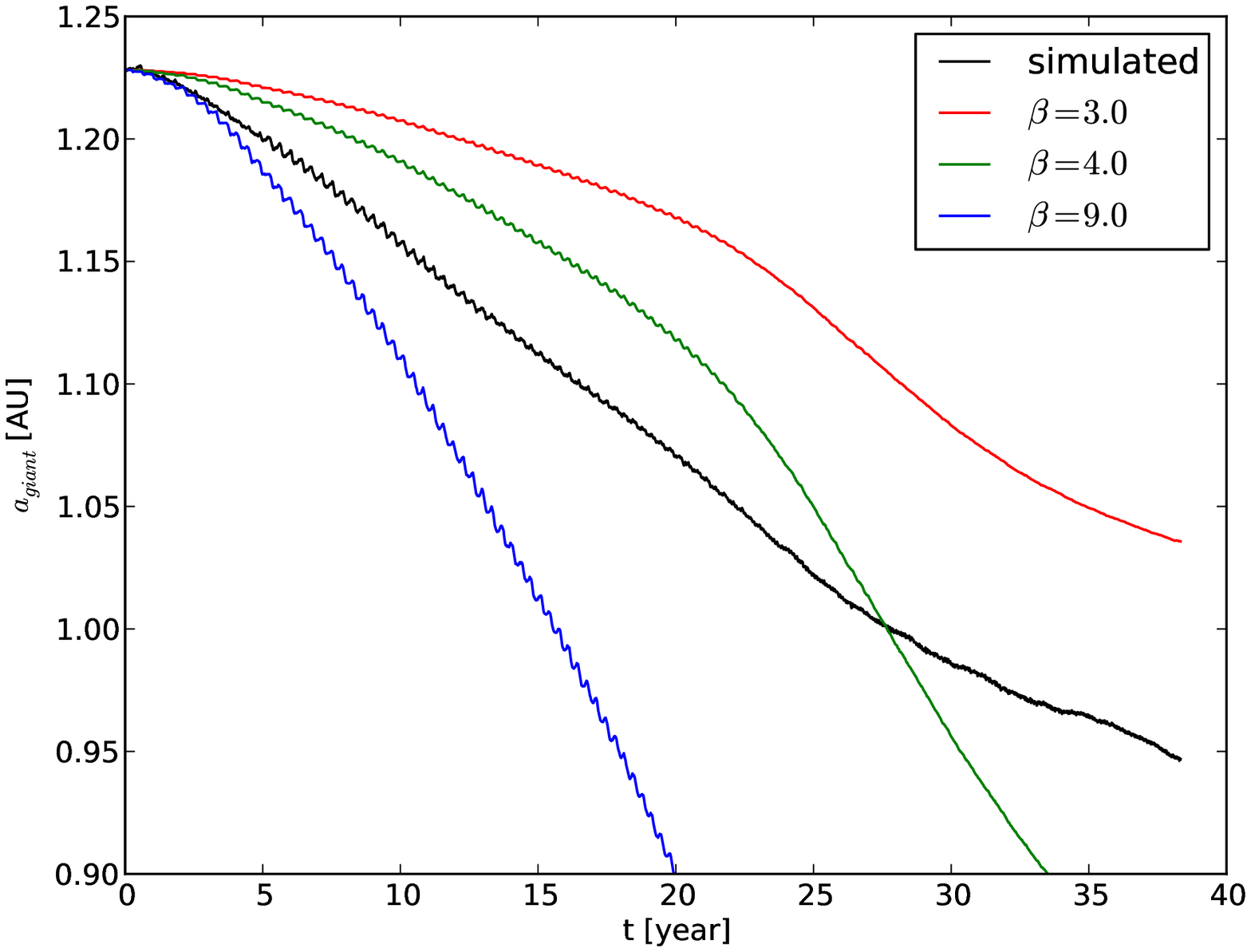,width=0.47\textwidth} 
 \psfig{file=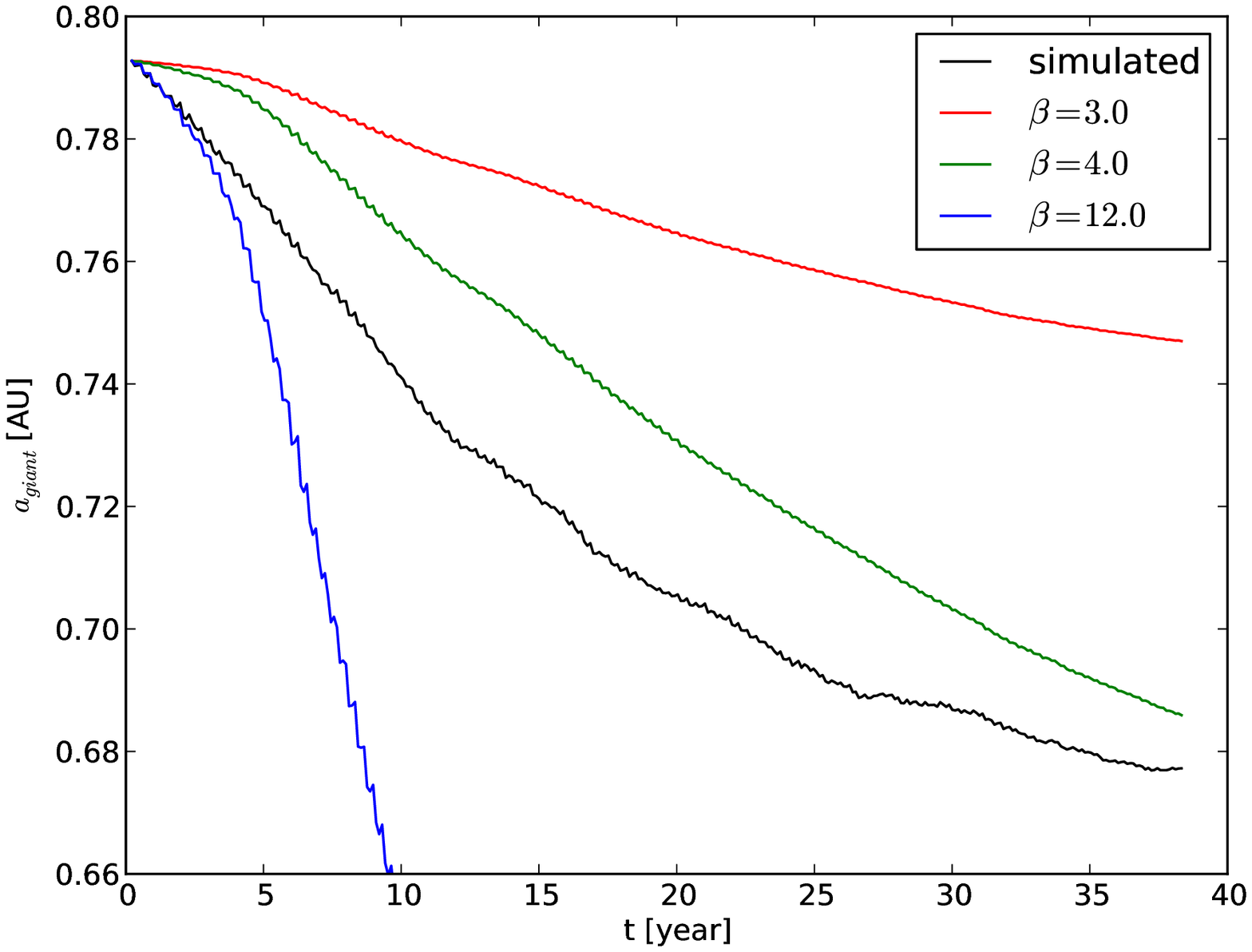,width=0.47\textwidth} 
 \caption{Semi-major
 axis as a function of time for \xitau\ (top panel) and \HD\ (bottom panel).  
 The black solid
 curve gives the actual calculated evolution using the simulations
 from \S\,\ref{Sect:CaseStudy}. The coloured curves give the orbital
 separation as a function of time using Eqn.\,\ref{Eq:NewSemi} for different values of $\beta$.
 \label{fig:OrbitalEvolution}
}
\end{figure}

\begin{figure}
 \centering 
 \psfig{file=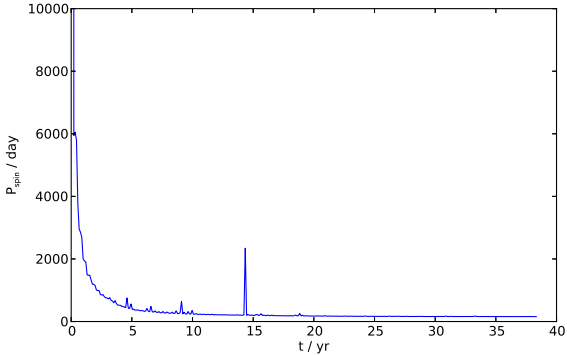,width=0.47\textwidth} 
 \caption{Evolution of the spin period of the outer star as a function of time for  \HD.  
 \label{fig:SpinEvolution}
}
\end{figure}

In Fig.\,\ref{fig:OrbitalEvolution} we present the orbital evolution
of \xitau\ and \HD\ (solid black curves). The coloured curves give the 
results of Eq.\ref{Eq:NewSemi} with comparable models, adopting values 
for $\beta$ of 3 (red curve), 4 (green curve), and 9 and 12 for 
\xitau\ and \HD, respectively (blue curve). The models with lower 
values for $\beta$ do not match the simulations very well in the 
beginning, but we argue that this is because the giant's orbit and spin 
are not yet in corotation. The time scale on which the giant gets into 
corotation is similar to the time scale at which the models with lower 
$\beta$ values start to match the slope of the orbital evolution, as 
can be seen by comparing Fig.\,\ref{fig:OrbitalEvolution} and 
Fig.\,\ref{fig:SpinEvolution}. After about 35 years the evolution of 
the semi-major axis of both \xitau\ and \HD\ match the model with 
$\beta\simeq3$.

\section{The future of \xitau\ and \HD}

\subsection{The future of \xitau}

\begin{figure}
 \centering
 \psfig{file=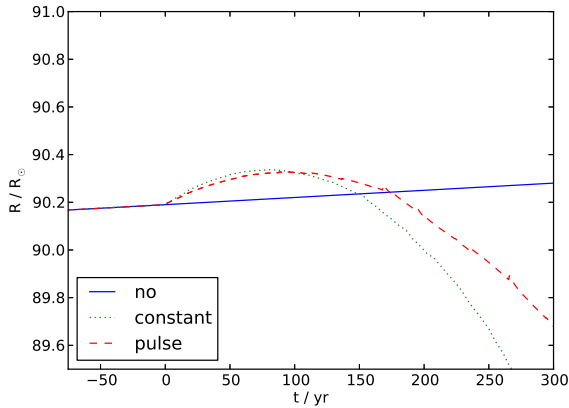,width=0.47\textwidth}
 \caption{Evolution of the size of the tertiary star in \xitau\
   starting from the moment of the RLOF phase ($t=0$). In the first
   millenium after the onset of RLOF the star grows quite dramatically
   to eventually drop.
}
 \label{fig:Radius_XiTau_RLOF}
\end{figure}

At the start of our hydrodynamics simulations the
outer star has a radius of $\sim 90\,\RSun$ and due to evolution
it continues to grow at a rate of about $0.0004\,\RSun$/yr.  With the
observed orbital parameters the size of the Roche lobe is
$\sim 82\,\RSun$ and $\sim 111\,\RSun$ at pericenter and apocenter, respectively, which means that the star at pericenter overfills
its Roche lobe but at apocenter is detached. The outer orbit will quickly circularize. By that time the separation decreased to 1.0\,AU for low initial inclination angles. The size of
the Roche lobe at that moment will be about $78\,\RSun$, and naively
speaking the star will continuously be filling its Roche lobe.  However, the
mass loss causes the giant to shrink at an accelerated rate and in
about 300\,years after the onset of mass transfer it detaches from its
Roche lobe.  In Fig.\,\ref{fig:Radius_XiTau_RLOF} we present the
evolution of the size of the outer star as calculated by \texttt{MESA} for different mass-loss schemes. The blue drawn line shows the radius evolution with no mass loss, for comparison. The green dotted line shows the radius evolution when the star loses mass at a constant rate. The red dashed line corresponds to the `pulsing' mass loss for eccentric orbits we see in our simulations, with no mass loss for 80\% of the orbit, and for 20\% of the orbit a mass loss five times higher compared to the constant mass loss scheme. We realize that when the radius drops significantly, the giant will detach from its Roche lobe and the radius evolution will continue the normal stellar evolution track (no mass loss), until RLOF is re-established. This will result in a radius evolution with a series of bumps like the one in Fig.\,\ref{fig:Radius_XiTau_RLOF}.

\begin{figure}
 \centering
 \psfig{file=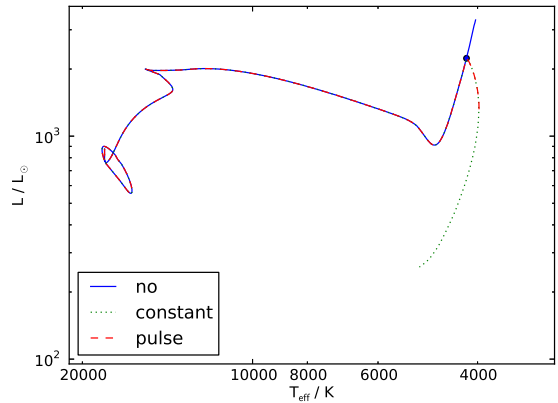,width=0.47\textwidth}
 \caption{Hertzsprung Russel diagram of the tertiary star in \xitau. 
   The evolution of the ZAMS star starts at the lower left
   corner, and the star evolves through a series of evolutionary
   phases until it reaches RLOF near the tip of the AGB (indicated
   with a bullet). During the RLOF phase the star loses mass in
   episodes which causes the evolution to be quenched, as indicated in
   the track to the bottom right after the bullet point.
  }
 \label{fig:HRD_XiTau}
\end{figure}

In Fig.\,\ref{fig:HRD_XiTau} we present the Hertzsprung-Russel diagram
for \xitau. The mass loss causes the outer star to drop in magnitude 
and become first red and later blue, below the giant branch but to the 
right of the main-sequence.  The extraordinary location of the star in
the Hertzsprung-Russel diagram would place it where some of the
sub-subgiant reside \citep{2003AJ....125..246M}. This position in the
Hertzsprung-Russel diagram is consistent with several sub-subgiants or
red stragglers\footnote{It is interesting to note that in each of the
  papers that classify ``red stragglers'' the authors make the claim
  that they coin the term. The earliest mention of the term that we
  could find was in \cite{1981A&A....99..221F}.}  found in NGC6791 (in
particular star \# V9, V17 and V76) \citep{2003AcA....53...51K}, which
in their interpretation are variable stars. Similar objects have also
been found in NGC6752 \citep{1997ApJ...474..701R}, 47Tuc
\citep{2001ApJ...559.1060A} and M67
\citep{2000AAS...197.4111M,2003AJ....125..246M} (S1036 and S1113). The
red stragglers in M67 and V9 in NGC6791 are confirmed binaries, and it
cannot be excluded that the others have an unseen binary companion.
The argument for the donor in the binary in our calculation to become 
a sub-subgiant is consistent with the explanation given earlier
\citep{1999A&A...347..866V}.

During the first half of the simulation, the outer orbit quickly circularizes and the giant's orbit and spin are driven into corotation. Subsequently, for low inclination angles, the rate of change of the inner binary semi-major axis steepens significantly, while the evolution of the outer orbit slightly flattens off (see Fig.\,\ref{fig:eXiTau} and Fig.\,\ref{fig:aXiTau}). 
For example for an initial inclination of $i=9^\circ$, the outer star loses $0.69\,\MSun$ (15\%) of its mass during the last 10\,years of the simulation, while the semi-major axes of the inner and outer orbits decreased by $1.3\,\RSun$ (4.9\%) and $10\,\RSun$ (4.7\%), respectively. Since the ratio of the fractional rates of change of the inner and outer semi-major axes is approximately unity,
\begin{displaymath}
  {(\dot{a}_{in} / a_{in}) \over (\dot{a}_{out} / a_{out})} \simeq 1,
\end{displaymath}
the triple remains dynamically stable. The system as a whole shrinks significantly during the RLOF phase, but not enough to initiate RLOF in the inner binary. 

After the outer star has turned into a white dwarf, the inner binary
will eventually acquire Roche-lobe contact and an entirely new phase in the evolution of the triple starts. 
We will not dwell on the further
evolution of the system, but the intermediate phase where a white
dwarf has a circular orbit around a close (also circularized) binary
could be interesting from an observational point of view.

\subsection{The future of \HD}

Roche-lobe overflow in \HD\ sets in around 3\,Gyr when the
$1.5\,\MSun$ donor has grown to a $56\,\RSun$ giant. 
Upon the initial Roche-lobe contact the accretion stream of the donor
arrives at a distance of $\sim 6.2\,\RSun$ from the center of mass
of the inner binary, i.e, it comes within the orbit of the inner
binary (which has an orbital separation of about $8\,\RSun$).  The
absence of an accretion disk in this system is then not a surprise.

Initially the outer orbit shrinks faster compared to the inner orbit, with a low ratio of the rates of change of the inner and outer semi-major axes 
\begin{displaymath}
  {(\dot{a}_{in} / a_{in}) \over (\dot{a}_{out} / a_{out})} \simeq 0.3.
\end{displaymath}
While circularizing, this ratio approaches unity, similar to the value found for \xitau.
In this case, however, the initial semi-major axis of the inner orbit is already very small, and the inner binary can possibly initiate RLOF. 

\section{Discussion and conclusions}

We performed a series of simulations using the \texttt{AMUSE} software
environment in order to study the gravitational dynamics,
hydrodynamics, and stellar evolution of hierarchical triple systems.
In the triples we selected the outer star is sufficiently massive and
the outer orbit sufficiently small for Roche-lobe overflow from the
outer star onto the inner binary to occur. These systems are
relatively rare $\aplt 1\%$ (6/725 in the catalogue), but 
the consequences of this RLOF phase can be profound.

We follow the onset of RLOF for two cases in detail for a few decades
using a combination of codes. We used a gravitational N-body code to
follow the dynamical evolution of the three stars, a smoothed-particles hydrodynamics code to study the accretion stream from the
outer star onto the inner binary system and a stellar evolution code
to continue the evolution of the outer Roche-lobe filling star.

In neither of the cases we studied an accretion disk formed, not even 
a circumbinary disk. The mass that leaves the outer 
star is funnelled through the first Lagrangian point of the outer 
orbit to land very near the inner binary system. The interaction 
between the gas and the inner binary causes the former to be ejected 
from the binary system like in a common envelope. {For systems with a 
high relative inclination, the evolution of 
the inner binary is adequately described by the common-envelope 
prescription with an efficiency parameter $\alpha\lambda \simeq 3$, 
which is consistent with the values derived for observed post-common-envelope binaries 
\citep{2012ApJ...759L..34C, 2013MNRAS.429L..45P}. 
Simulations with lower relative inclination result in higher values for $\alpha \lambda$, which 
is expected since the angular momentum vectors of the inner binary and the inflowing gas are aligned, and less angular momentum needs to be transferred to speed up the gas to the escape velocity.} For the mass loss
from the outer binary system we measure a value of $\beta \simeq 3$ to $4$,
which means that the mass lost from the binary carries less angular
momentum per unit mass compared to what is available in the second Lagrangian
point. However, also here the values of both systems studied in detail
give a consistent result.

The rapid mass loss causes the outer giant to shrink and it may even
detach from its Roche lobe, which temporarily would interrupt the mass
transfer. The orbital separation of the inner binary therefore shrinks
in time until the inner binary will itself engage in a phase of RLOF
or the envelope of the outer star is depleted.

In one of our studied cases, \xitau, the mass transfer causes the inner and outer orbits to shrink at the same fractional rate, 
\begin{displaymath}
  {(\dot{a}_{in} / a_{in}) \over (\dot{a}_{out} / a_{out})} \simeq 1,
\end{displaymath}
and this system is therefore likely to remain dynamically
stable while the entire envelope of the outer star is transferred
to the inner binary. 
After the main-sequence phase of the inner binary stars, a common-envelope 
phase with all three stars embedded is likely to
result in a collision between all three stars, although we did not
explicitly test this hypothesis.

In the other case, of \HD, the ratio of the rates of change of the inner and outer semi-major axes also approaches unity, and initially the
triple is dynamically stable as well. However, the inner binary is much more compact in this system, and the decrease in semi-major axis can possibly lead to simultaneous RLOF in the inner binary.

In both cases the detailed evolution, for as long as we followed it
hydrodynamically, resulted in systems that have interesting
observational implications. The phase we followed, however, is very
short. The
chance of observing the mass transfer in action is therefore rather
small. The long term effect, however, are more likely to be
observable. A merger between a giant and two main-sequence stars
could lead to some sort of curious stellar object with non-canonical
atmospheric and spectral features, but also the white dwarf in a
circular orbit around a tight binary would be an interesting finding.

Although our theoretical analysis is limited in scope and detail,
we would like to point out, and we are not to first to say this, that 
the evolution of triples is quite complicated, and not a simple 
extension of binary evolution with some third component. The mutual 
interaction in terms of hydrodynamics, stellar evolution and the 
dynamical complications make for a challenging topic to study, and 
very hard to get a general consensus on triple evolution.

{\bf Acknowledgements} It is a pleasure to thank 
Adrian Hamers, Edward P.J.\,van den Heuvel, Inti Pelupessy, Arjen
van Elteren, and the anonymous referee for comments on the manuscript and discussions.  This work
was supported by the Netherlands Research Council NWO (grants
\#612.071.305 [LGM], \#639.073.803 [VICI] and \#614.061.608 [AMUSE])
and by the Netherlands Research School for Astronomy (NOVA). We made use of pynbody (http://code.google.com/p/pynbody) for generating the column density images (Fig.\,\ref{fig:StillsXiTau}) in this paper.

\bibliographystyle{mn2e}
\bibliography{MN-13-1114-MJ.R1}

\end{document}